\documentclass[a4paper, fleqn, oneside, 10pt]{article}

\usepackage{amsmath}
\usepackage{amsfonts}
\usepackage{amsthm}
\usepackage{verbatim}

\DeclareMathOperator{\supp}{supp}
\DeclareMathOperator{\diverg}{div}

\DeclareMathOperator{\im}{Im}
\DeclareMathOperator{\tr}{Tr}
\DeclareMathOperator{\I}{I}
\DeclareMathOperator{\ran}{ran}
\DeclareMathOperator{\rea}{Re}

\newtheorem{lemma}{Lemma}[section]
\newtheorem{theorem}[lemma]{Theorem}
\newtheorem{proposition}[lemma]{Proposition}

\newtheorem{corollary}[lemma]{Corollary}
\theoremstyle{definition}
\newtheorem{definition}[lemma]{Definition}
\theoremstyle{definition}
\newtheorem{remark}[lemma]{Remark}
\theoremstyle{definition}
\newtheorem{example}[lemma]{Example}

{\catcode `\@=11 \global\let\AddToReset=\@addtoreset}
\AddToReset{equation}{section}
\renewcommand{\theequation}{\thesection.\arabic{equation}}

\newcommand{\ie}{{\sl i.e.\/ }}
\newcommand{\cf}{{\sl cf.\/ }}
\newcommand{\eg}{{\sl e.g.\/ }}

\newcommand{\newpar}{\par}\parindent =0pt\parskip=3pt\textheight = 615pt
\renewcommand{\L}{{L}}
\newcommand{\Id}[1]{{\rm I\kern-2pt I_{#1}}}
\renewcommand{\hbar}{{\displaystyle\bar{\phantom{x}}\kern-6pt h}}
\newcommand{\fer}[1]{(\ref{#1})}
\newcommand{\astx}{{}_{\stackrel{\displaystyle{\ast}}{x}}}
\newcommand{\asty}{{}_{\stackrel{\displaystyle{\ast}}{y}}}
\def\3norm{|\!|\!|}

\begin{document}

\thispagestyle{empty}
\begin{center}
\fontsize{20}{22}
\selectfont
\textbf{Quantum dynamical semigroups for diffusion models with Hartree interaction}
\end{center}
\vskip 1 cm
\begin{center}
\fontsize{12}{12}
\selectfont
A. Arnold\footnote{Institut f\"ur Numerische Mathematik,
Universit\"at M\"unster,
Einsteinstr. 62,
D-48149 M\"unster, Germany,
e-mail: anton.arnold@math.uni-muenster.de,} and
C. Sparber\footnote{Institut f\"ur Mathematik, Universit\"at
Wien, Strudlhofgasse 4, A-1090 Vienna,
Austria,\\
e-mail: christof.sparber@univie.ac.at.}
\end{center}
\vskip 1 cm
\begin{center}
\textbf{Abstract}
\end{center}
\emph{We consider a class of evolution equations in Lindblad form,
which model the dynamics of dissipative quantum mechanical systems
with mean-field interaction. Particularly, this class includes the
so-called Quantum Fokker-Planck-Poisson model. The existence and
uniqueness of global-in-time, mass preserving solutions is proved, thus
establishing the existence of a nonlinear conservative quantum
dynamical semigroup. The mathematical difficulties stem from combining an unbounded
Lindblad generator with the Hartree nonlinearity.} \vskip 1cm \noindent \textbf{Key words:}
open quantum system, Lindblad operators, quantum dynamical
semigroup, dissipative operators, density matrix, Hartree equation
\newpar

\vskip 1cm


\section{Introduction}

This paper is concerned with quantum mechanical multi-particle systems coupled to an
external reservoir, \ie so called \emph{open quantum systems} \cite{Da, BrPe}. The
dynamics of such systems can often be approximately described by kinetic equations in
the \emph{mean-field limit}.
Such \emph{self-consistent models} appear in a wide range of physical applications, both
quantum mechanical and classical, for example in gas dynamics, stellar dynamics, plasma physics,
and electron transport. The corresponding nonlinear evolution equations are obtained as
approximations to the underlying (linear) many-particle models, and there exists
a vast body of
literature on their mathematically rigorous derivation: the classical
\emph{Vlasov-Poisson} system in \cite{BrHe, Ba}; the \emph{Hartree equation}  from the 
$N$-body Schr\"odinger equation in the \emph{mean-field limit} in \cite{ErYa};
the \emph{Hartree-Fock equation} in \cite{BaMa}. All of these models have in common that they fall 
into the class of \emph{Markovian approximation} for the underlying dynamics and we refer to \cite{Sp} 
for an extended overview of such derivations for a variety of kinetic equations. 
\newpar
In addition to a self-consistent Coulomb field we shall here be interested in
quantum systems which in addition have a \emph{dissipative} interaction with their 
environment. 
In many (practical) applications of such open quantum systems the interaction with a reservoir is
described in a rather simple phenomenological manner, often using diffusion operators,
quantum-BGK or relaxation-type terms \cite{CaLe, DeRi, Ar1} when considered in a
kinetic formalism. A prominent example of a \emph{linear} open quantum
system is the so called \emph{quantum optical master equation} and its variants \cite{GaZo, Va1}. 
However \emph{nonlinear} mean-field models for open quantum systems also play an important role \eg in laser
physics (\cf \cite{HeLi} and \cite{Sp} for the \emph{Lieb-Hepp} and the 
\emph{Dicke-Haken-Lax laser model}, resp). 
\newpar
In this work we shall be interested in a particular class of models which are frequently used in 
quantum optics \cite{DHR, OC, Va1} and the simulation of nano-scale semiconductor devices \cite{FMR, JuTa}, 
namely the quantum kinetic \emph{Wigner-Fokker-Planck equation} (WFP) 
\begin{equation}\label{wfp}
\partial_t w +\xi \cdot\nabla _x w +\Theta [V] w =  \mathcal Qw,
\quad\quad x,\xi \in\mathbb R^d, t>0,
\end{equation}
which governs the time evolution of the Wigner function $w(x,\xi,t)$ in 
(position-velocity) \emph{phase-space} under the action of the potential $V(x,t)$.
In \fer{wfp}  the pseudo-differential operator $\Theta [V]$ is defined by
\begin{align}
\Theta [V]w(x,\xi,t) := \frac{i}{(2\pi )^d}\int\int_{\mathbb R^d\times\mathbb R^d}
\left[V\left({ x+\frac{y }{2},t}\right)-V\left({ x-\frac{y}{2},t}\right)\right]
\nonumber\\  w(x,\xi',t)\ e^{iy\cdot (\xi
-\xi')}d\xi 'dy.
\end{align}
$\mathcal Q$ denotes the following diffusion operator 
\begin{equation}
\label{lq}\mathcal Qw(x,\xi):=\ D_{pp} \Delta _\xi w
+2\eta \diverg_\xi (\xi w ) + D_{qq} \Delta _xw
+2 D_{pq} \diverg_x(\nabla _\xi w ),
\end{equation}
with diffusion constants $D$ (cf. \fer{li} below) and the friction constant $\eta\ge0$.
Here and in the sequel we set the physical constants $\hbar=m=e=1$, for
simplicity. In semiconductor applications $w(t,x,\xi)$ is the quasi-distribution of the electron gas
and $\mathcal Q$ models (phenomenologically) its interaction with a phonon bath. In our
mean-field model the \emph{Hartree-type nonlinearity} then stems from the repulsive Coulomb
interaction between the electrons. Hence, \fer{wfp} is coupled to the \emph{Poisson equation}
\begin{equation}\label{poisson}
\Delta V  = -n,
\end{equation}
where $n=\int w\,d\xi$ is the \emph{particle density} of the electrons.
\newpar
Moreover, such  \emph{Quantum Fokker-Planck} (QFP) type equation
are the most prominent model in the  description of \emph{quantum
Brownian motion}, where a (massive) quantum particle interacts with a heath 
bath and a possible external potential, see \eg \cite{CaLe, De, Di, Li1, OC}
and \cite{HuMa},  where this setting is proposed as a description of
\emph{decoherence}.  Indeed most of these equations can be traced back to an
early work by Feynman and Vernon \cite{FeVe}. While formal derivations of QFP
equations were given in \cite{CaLe, Di1, Va}, a rigorous derivation from many-body
quantum mechanics is still missing,  at least for the general class of models
considered here. To the authors' knowledge, the only results in this direction
are  \cite{CEFM, FMR}, where special cases of the QFP equation arise, resp., in a 
\emph{space-time scaling limit} and a \emph{weak coupling limit}  for a
particle interacting with an infinite heat bath of harmonic oscillators, \ie 
\emph{phonons}. 
\newpar
In this paper we shall investigate well-posedness of QFP type equations with a
mean-field Coulomb potential -- the above mentioned \emph{Wigner-Poisson-Fokker-Planck equation}
(WPFP) \fer{wfp}-\fer{poisson} being one typical example. Specifically, we 
establish  existence and uniqueness of global-in-time solutions to the Cauchy problem.
Many of the analytical tools developed in the sequel will, however, directly apply to
other open quantum systems in mean-field approximation 
(\eg to the Dicke-Haken-Lax laser model).
First analytical results on the WFP and WPFP
equations (\ref{wfp}) were obtained in \cite{SCDM} (well-posedness of the linear
equation, convergence to the unique steady state with an exponential rate), in
\cite{ALMS} (local-in-time solution for the mean-field model in 3D), and in
\cite{ACD} (global-in-time solution for the mean-field model in 1D). 
\newpar
In the mathematical analysis of mean-field QFP equations several parallel problems have
to be coped with: the Wigner framework often used in applications seems inappropriate since the particle density
$n=\int w\,d\xi$ is not naturally defined in this setup (typically, $w\in L^2(\mathbb
R^d_x\times \mathbb R^d_\xi)$;  \cf \cite{Ar, ALMS} for more details). We are hence led
to study the equivalent evolution of the \emph{density matrix} $\rho(t)$ in the space of 
positive trace class operators $\mathcal J_1$. Moreover, in order to deal with the Hartree nonlinearity, 
an appropriate energy-space $\mathcal E\subset \mathcal J_1$ needs to be introduced, 
which is a generalization of the one used in \cite{BDF}.
In $\mathcal J_1$ the evolution of the quantum system is then governed by a so called \emph{Markovian master equation},
\begin{equation*}
\left \{
\begin{split}
&\frac{d}{dt} \rho = \mathcal L (\rho) , \quad t>0,\\
&\rho\big |_{t=0}=\rho_0\in \mathcal J_1.
\end{split}
\right.
\end{equation*}
The considered \emph{Liouvillian} $\mathcal L$ is obtained as a generalization of the one given by an inverse Wigner transformation of 
\fer{wfp} and will be stated in \fer{maeq} below. 
Since $\mathcal L$ (and in particular  the included Lindblad operators
\cite{Li}) are unbounded, this can be difficult even for linear equations and
may lead to \emph{non unique} and \emph{non conservative solutions}. E.B.
Davies showed in \cite{Da1} that it is possible to construct, for a quite
general class of unbounded Lindblad generators $\mathcal L$, a so called
\emph{minimal solution} to the above master equation. However, this
construction is in general not unique, \ie $\mathcal L$ does not uniquely
determine a corresponding \emph{quantum dynamical semigroup} (QDS) $\Phi
_t(\rho_0) =e^{\mathcal Lt}\rho_0$. In particular, this implies that the
minimal solution may not be \emph{conservative}, \ie trace preserving (\cf
example 3.3 in \cite{Da1}), which would be  inappropriate  for the above mentioned applications.
\newpar
While linear QDS have been studied intensively in the last three decades \cite{FaRe, Al,
AlFa}, the literature on nonlinear QDS is no so abundant, see \eg \cite{Ar1, AlMe, BDF}.
By now, various sufficient conditions for the conservativity of linear  QDS can be found
in \cite{ChFa, CGQ, Ho}. For many concrete examples, however, these conditions are
rather difficult to verify, as we shall discuss in more detail at the end of section
\ref{lins}. Moreover the assumptions on the nonlinearity introduced in \cite{AlMe} seem too strong 
for most physical applications.
\newpar
In this perspective, the present work establishes the existence and uniqueness of a conservative QDS for a
concrete family of unbounded Lindblad generators $\mathcal L$ (including the WPFP model) with Hartree interaction.
We shall consider Lindblad operators (representing the coupling to the reservoir) which 
are linear combinations of the position and momentum operators, \ie so called 
\emph{quasifree dynamical semigroups} \cite{Li1}. 
\newpar
We briefly remark that the classical counterpart of WPFP, \ie the 
\emph{Vlasov-Poisson-Fokker-Planck} system (and its linear version, the 
classical kinetic Fokker-Planck or \emph{Kramers equation} \cite{Ri})
\begin{equation}\label{vfp}
\partial_tf +\xi \cdot\nabla_xf -\nabla_x V\cdot\nabla_\xi f 
= D\Delta_\xi f+2\eta \diverg_\xi (\xi f ),\quad\quad x,\xi \in\mathbb R^d, t>0
\end{equation}
allows for a much easier mathematical analysis. This is due to a natural $L^1(\mathbb R^d_x\times \mathbb R^d_\xi)$--framework
for \fer{vfp} and to the positivity of the phase-space density $f(t,x,\xi)$, \cf \cite{Bo} for the
well-posedness analysis, \cite{Dr} for existence of a unique steady state, and \cite{DeVi} for
convergence results to the steady state for the linear model. 


\newpar
This paper is organized as follows:
\newpar
After introducing the model in section \ref{sets} we will prove in
section \ref{lins} existence and uniqueness of a global, mass
preserving solution to the
linear equation, \ie the existence of a conservative QDS. A crucial analytical tool towards this end is a new
density lemma (relating minimal and maximal
operator realizations) for Lindblad generators $\mathcal L$ that are quadratic in
the position and momentum operator. 
The mean field will then be included in section
\ref{nlinl} (we shall restrict ourselves for simplicity to the case of $d=3$ spatial dimensions).
We prove that the self-consistent potential is a locally Lipschitz perturbation of the free evolution in an
appropriate ``energy space'', and this yields a local-in-time existence and uniqueness
result.
Finally, we shall prove global existence of a conservative
QDS in section \ref{nling} by establishing a-priori estimates for the mass 
and total energy of the system.


\section{The model equation} \label{sets}

In the sequel we shall use the following standard notations:
\begin{definition} 

$\mathcal J_1$ is the space of \emph{trace class operators} on  $L^2(\mathbb
R^d)$ with the norm ${\3norm A\3norm}_1 := \tr |A|$, where $\tr$ denotes the
usual operator trace on $\mathcal B(L^2(\mathbb R^d))$. $\mathcal J_1^s
\subset \mathcal J_1$ denotes the subspace of \emph{self-adjoint} trace
class operators. Similarly, $\mathcal J_2$ is the space of
\emph{Hilbert-Schmidt operators} with the norm ${\3norm A\3norm}_2 :=
\left(\tr|A|^2\right)^{1/2}$ and $\3norm \cdot \3norm_\infty$ denotes
the \emph{operator norm} in  $\mathcal B(L^2(\mathbb R^d))$. ${\| \cdot
\|}_p$, $1\leq p\leq \infty$ is the norm of $L^p(\mathbb R^d)$-functions.

\end{definition}
We consider open quantum systems of massive, \emph{spin-less} particles within an
\emph{effective single-particle approximation}, as it has been derived for example in \cite{CEFM}.
Hence, at every time $t\in\mathbb R$ a physically relevant, \emph{mixed state}
of our system is uniquely given by a positive operator $\rho(t)\in \mathcal J_1^s$, in the sequel called
\emph{density matrix operator}. Since $\rho$ is also
Hilbert-Schmidt it can be represented by an integral operator
$\rho(t): L^2(\mathbb R^d)\rightarrow L^2(\mathbb R^d)$, \ie
\begin{equation}
\label{dop} (\rho(t)f)(x):=\int_{\mathbb R^d} \rho (x,y,t)f(y)dy.
\end{equation}
Its kernel $ \rho (\cdot,\cdot,t  )\in L^2(\mathbb R^{2d})$ is
then called the \emph{density matrix function} of the state $ \rho$ 
and it satisfies ${\3norm\rho(t)\3norm}_2 ={ \| \rho(\cdot, \cdot, t) \|}_2$. 
By abuse of notation we shall identify from now on the operator
$ \rho \in \mathcal J_1^s$ with its kernel $ \rho (\cdot ,\cdot )\in
L^2(\mathbb R^{2d})$.  
It is well known that we can decompose the kernel in the following form
\begin{equation}
\label{fek}\rho(x,y)=\sum_{j\in\mathbb N} \lambda _j \ \psi _j(x) \overline{\psi _j(y)},
\quad \lambda _j\geq 0,
\end{equation}
where $\{\lambda _j\}\in {\it l}^1(\mathbb N)$ and the complete 
o.n.s. $\{\psi _j\} \subset L^2(\mathbb R^d)$ are
the eigenvalues and eigenfunctions of $\rho$. 
Using equation (\ref{fek}) one can define the \emph{particle
density} $n[\rho]$ by setting $x=y$, to obtain
\begin{equation}
n[\rho](x):=\sum_{j\in\mathbb N} \lambda _j \ |\psi _j(x)|^2,\quad x \in
\mathbb R^d.
\end{equation}
However, since $\{x=y\}\subset \mathbb R^{2d}$ is a set of measure
zero, this is not a mathematically rigorous procedure for a kernel
$\rho(x,y)$ that is merely in $L^2(\mathbb{R}^{2d})$. On the other
hand, if $\rho (x,y)$ is indeed the kernel of an operator $\rho
\in \mathcal J_1$ it is known, \cf \cite{Ar}, \cite{LiPa}, that
the particle density can be rigorously defined by
\begin{equation}\label{den}
n[\rho](x):= \lim_{\varepsilon \rightarrow 0} \int_{\mathbb R^d} \rho
\left(x+\frac{\eta }{2}, x-\frac{\eta }{2}\right)
\frac{e^{-|\eta |^2/2\varepsilon }}{(2\pi \varepsilon )^{d/2}} \ d\eta \in L^1_+(\mathbb R^{d}).
\end{equation}
And it satisfies $\|n\|_1 = \tr(\rho)$ for $\rho \ge 0$. This
issue of rigorously defining $n[\rho]$ is one of the mathematical
motivations for analyzing our mean field evolution equations as an
abstract evolution problem for the operator $\rho$ on the Banach space $\mathcal J_1^s$ 
.
\begin{remark}
Note that we can not use the decomposition (\ref{fek}) in order to pass to
a PDE problem for the $\psi_j$, since the considered dissipative evolution equation in general does
not conserve the occupation probabilities $\lambda_j$. This is in sharp contrast to
unitary dynamical maps generated by the von Neumann equation of standard quantum mechanics.
\end{remark}
We consider the following (nonlinear) dissipative equation modeling the motion of
particles, interacting with each other and with their environment
\begin{equation}\label{maeq}
\left \{
\begin{split}
&\frac{d}{dt} \rho = \mathcal L (\rho):=-i \left[H, \rho\right] + A(\rho), \quad t>0,\\
&\rho\big|_{t=0}=\rho_0\in \mathcal J_1^s.
\end{split}
\right.
\end{equation}
Here, $[\cdot ,\cdot ]$ is the commutator bracket, $H$ and
$A(\rho)$ are formally self-adjoint and of Lindblad class. More
precisely, we consider the \emph{Hamiltonian operator}
\begin{equation}\label{ham}
H := -\frac{\Delta}{2}  +V[\rho](x,t) - i \mu {[x, \nabla]}_+, \quad \mu \in \mathbb R,
\end{equation}
denoting by ${[\cdot ,\cdot ]}_+$ the anti-commutator. The operators $x$ and
$\nabla$ are, respectively, the multiplication and
gradient operator on $\mathbb R^d$, \ie ${[x, \nabla]}_+=x\cdot \nabla +\nabla \cdot x = 2x \cdot \nabla + d$.
\begin{remark} The operator $H$ is sometimes called \emph{adjusted Hamiltonian}, due to the appearence of the
$[x,\nabla]_+$- term. Depending on the particular model, such a
term may or may not be present, see \eg \cite{De, Di1}. Nevertheless
it is included here, in order to keep our presentation as general
as possible.
\end{remark}
The (real-valued) potential $V$ is assumed to be of the form
\begin{equation}
\label{pot} V[\rho](x,t):= \frac{|x|^2}{2} +
V_1(x)+\phi [\rho](x,t),\quad x \in \mathbb R^d,
\end{equation}
where  the first term of the r.h.s. denotes a possible confinement
potential and $V_1 \in L^\infty (\mathbb R^d)$ is a bounded
perturbation of it. We point out that the quadratic confinement potential is
\emph{not} necessary for the subsequent mathematical analysis, it is just an
option. $\phi $ is the \emph{Hartree-} or \emph{mean field-potential},
obtained from the \emph{self-consistent} coupling to the \emph{Poisson equation}
\begin{equation}\label{pois}
-\Delta \phi [\rho]  = n[\rho].
\end{equation}
For $d=3$, we therefore get the usual Hartree-term:
\begin{equation}
\label{spot}
\phi [\rho](x,t) = \frac{1}{4\pi  } \int_{\mathbb R^3}
\frac{n[\rho](y,t)}{|x-y|} \ dy,\quad x,y\in \mathbb R^3,
\end{equation}
where $n$ is computed from $\rho $ by (\ref{den}). This
mean field approximation describes the (repulsive) Coulombian interaction of the particles
with each other.
\newpar
The non-Hamiltonian part is
defined as
\begin{align}
\label{lind} A(\rho):= &\ \sum_{j=1}^{m} L_j \rho L_j^*-\frac{1}{2}\left[L_j^*L_j,
\rho \right]_+,\quad m\in \mathbb N,
\end{align}
or equivalently
\begin{align}
A(\rho)= & \ \sum_{j=1}^{m} \frac{1}{2} \left[L_j \rho, L_j^*\right]+ \frac{1}{2}\left[L_j,\rho
L_j^*\right],
\end{align}
where the linear operators  $L_j$ (\emph{Lindblad operators}) are assumed to be of the form
\begin{align}
\label{lind1}L_j := \alpha_j \cdot x + \beta_j \cdot \nabla+ \gamma_j,
\quad \alpha_j ,\beta_j \in\mathbb C^d, \gamma_j\in \mathbb C.
\end{align}
Its adjoint is $L^*_j = \bar \alpha_j\cdot x-\bar \beta_j \cdot \nabla +\bar \gamma_j $, and
in the following we shall use the notation
\begin{equation}
L:=\sum_{j=1}^m L_j^*L_j.
\end{equation}
\begin{remark} \emph{Linear} models with Hamiltonians that are quadratic in the position
and momentum operator and with Lindblad operators of the form \fer{lind1} give
rise to so called \emph{quasifree QDS}, and they are explicitly solvable in
terms of Greens functions \cite{Li1, SCDM}. In oder to deal with nonlinear
problems (in a ``finite energy subspace'' of ${\mathcal J_1}$) we shall, however,
not use this representation, which moreover can not be generalized to higher order models, \cf 
remark \ref{JC}.
\end{remark} 
\begin{remark} In the framework of \emph{second quantization} and in $d=1$, the space $L^2(\mathbb R)$
is unitarily mapped onto $\mathcal F_s(\mathbb C)$, the \emph{symmetric} or \emph{bosonic Fock space}
over $\mathbb C$.
This space is frequently used, for example in quantum optics, in order to describe $two-level$
bosonic systems,  \cf \cite{AlFa}, \cite{GaZo}. \\
Assuming $\mathbb \gamma =0$, $\beta=1$ and $\alpha=1/2$, the Lindblad operators $L$, $L^*$,
become then the usual bosonic \emph{creation-} and \emph{annihilation-operators}
\begin{align}
af(x):=(\frac{x}{2}+ \partial_x )f(x),\quad a^*f(x):=(\frac{x}{2} - \partial_x )f(x),
\end{align}
which, in contrast to the corresponding \emph{fermionic} creation- and anihilation-operators, are unbounded.
Of course, all results in our work can be equivalently interpeted in this framework of second quantization.
\end{remark}
\newpar
\begin{example} \label{ex.2.5} A particularly interesting example in the above class
is the \emph{Quantum Fokker-Planck equation} (QFP). As a PDE for the kernel $\rho(t,x,y)\in
L^2(\mathbb R^{2d})$ it reads
\begin{equation}
\label{exqfp}
\left \{
\begin{split}
i\partial_t\rho = \ & \frac{1}{\hbar }\left[-\frac{\Delta}{2}+V(t,x), \rho \right]+i A(\rho),\quad t>0, \\
\rho \big|_{t=0}= \ & \rho _0(x,y)\in L^2(\mathbb R^d_x\times \mathbb R^d_y), 
\end{split}
\right.
\end{equation}
where 
\begin{equation}
\begin{split}
A(\rho):= \, & -\gamma (x-y)\cdot (\nabla _x-\nabla _y)\, \rho +D_{qq} |\nabla _x+\nabla _y|^2 \rho \\ 
& -\frac{D_{pp}}{\hbar ^2}|x-y|^2\rho +  \frac{2iD_{pq}}{\hbar }(x-y)\cdot (\nabla _x+\nabla _y)\rho .
\end{split}
\end{equation}
This model can be written in the form (\ref{maeq}), (\ref{lind}), iff
the conditions
\begin{equation}
\label{li} D_{pp}D_{qq}-D_{pq}^2 \geq\frac{\eta ^2}{4}, \quad D_{pp},D_{qq}\geq 0,
\end{equation}
hold (see \cite{Li1, ALMS} for more details and a particular choice of
the parameters $\mu$, $\alpha_j$, $\beta_j$, $\gamma_j$). 
Using the \emph{Wigner transform} \cite{Wi, LiPa}: 
\begin{equation}
w (x,\xi ,t):=\frac{1}{(2\pi )^d}\int_{\mathbb R^d}\rho
\left( x+\frac{y}{2}, \ x-\frac{y}{2},
t\right)e^{i\xi \cdot y}dy.
\end{equation}
the QFP equation (\ref{exqfp}) can be 
transformed into the kinetic Wigner-Fokker-Planck equation \fer{wfp}.
In physical units  $D_{qq}, D_{pq}\sim O(\hbar^2)$, \cf \cite{De, Va}, and
hence we indeed  obtain, at least formally, the kinetic Fokker-Planck equation  
\fer{vfp} in the  (semi-)classical limit $\hbar\rightarrow 0$. Note that for
$\eta >0$, condition (\ref{li}) implies that the diffusion operator $\mathcal
Q$ from (\ref{lq}) is uniformly elliptic, which disqualifies the classical FP
diffusion operator (i.e.~$D_{qq} = D_{pq}= 0$) \cite{Ri} as an appropriate quantum
mechanical equation. Nevertheless, this \emph{Caldeira-Leggett} master equation
\cite{CaLe} is sometimes used in applications as a phenomenological quantum
model, \cf \cite{St}.   
\end{example}

\begin{remark} \label{JC}

To close this section we mention an interesting model from quantum optics which
is not yet covered by our present analysis.
The \emph{Jaynes-Cumming model with phase damping} reads 
\begin{equation}
\frac{d}{dt} \rho = -i \left[H, \rho\right] + \kappa [H[H,\rho]],
\end{equation}
where $\kappa \in \mathbb R_+$ denotes the damping constant, \cf \cite{Lo}.
Since it involves Lindblad operators $L_j$ that are quadratic polynomials of the
position and momentum operators, it will be the focus of future research to
(hopefully) extend the lemma  \ref{le} (below) to such cases. 
\end{remark}


\section {Existence of a conservative QDS for the linear problem} \label{lins}

\newpar
We consider the linear evolution problem on $\mathcal J_1^s(L^2(\mathbb R^d))$
\begin{equation}\label{lqds}
\left \{
\begin{split}
&\frac{d}{dt} \rho = \mathcal L (\rho) , \quad t>0,\\
&\rho\big |_{t=0}=\rho_0\in \mathcal J_1.
\end{split}
\right.
\end{equation}
Here, $\mathcal L(\rho):=-i \left[H, \rho\right] + A(\rho)$ is the formal generator of a QDS on
$\mathcal J_1^s$, with
\begin{equation}
H=-\frac{\Delta}{2}+ \frac{|x|^2}{2} +V_1(x)-i \mu[x,\nabla]_+.
\end{equation}
\begin{definition} \label{def}
Given any Hilbert space $\mathcal H$,
one defines a \emph{conservative quantum dynamical semigroup} (QDS) as a
one parameter $C_0$ - semigroup of bounded operators
\begin{equation}
\Phi_t: \mathcal J_1(\mathcal H)\rightarrow  \mathcal J_1(\mathcal H),
\end{equation}
which in addition satisfies:
\newpar
\textbf{(a)} The dual map $\Phi^*_t :\mathcal B(\mathcal H) \rightarrow  \mathcal B(\mathcal H)$,
defined by
\begin{equation}
\tr (A \Phi_t(\rho ))=\tr(\Phi _t^*(A)\rho) ,
\end{equation}
for all $\rho \in \mathcal J_1(\mathcal H)$, $A\in
\mathcal B(\mathcal H)$,  is \emph{completely positive}. This
means that the map
\begin{align}
\Phi^*_t \otimes \I_n : \mathcal B(\mathcal H) \otimes \mathcal
B(\mathcal H_n) \rightarrow \mathcal B(\mathcal H) \otimes
\mathcal B(\mathcal H_n)
\end{align}
is positive (\ie positivity preserving) for all $n\in \mathbb N$.
Here $\mathcal H_n$ denotes a finite dimensional Hilbert space and $\I_n$ is the $n-dimensional$ unit matrix.
\newpar
\textbf{(b)} $\Phi_t$ is trace preserving, \ie conservative (or unital).
\end{definition}
\newpar
\begin{remark} The notion QDS is sometimes reserved for the dual
semigroup $\Phi _t^*$. Physically speaking, this corresponds to
the \emph{Heisenberg picture}. The appropriate continuity is then
\begin{equation}
\lim_{t\rightarrow 0} \tr (\rho(\Phi_t^*(A)-A)) =0,
\end{equation}
for all $\rho \in\mathcal J_1(\mathcal H) $, $A \in \mathcal B(\mathcal H) $, \ie \emph{ultraweak continuity}.
Complete positivity can be defined also for operators on general $C^*$-Algebras $\mathcal A$ \cite{Sti} and it is known that
complete positivity and positivity are equivalent only if $\mathcal A$ is commutative.
(Counter-examples can be found already for $2\times 2$ complex valued matrices, see \eg \ \cite{AlFa}.)
Again, from a physical point of view, complete positivity can be interpreted as preservation
of positivity under \emph{entanglement}.
\end{remark}
Following the classical work of Davies \cite{Da1} we shall start to investigate the
properties of the operator
\begin{equation}
\label{y} Y:= -iH-\frac{1}{2} L.
\end{equation}
First we need the following technical lemma, the proof of which introduces some important notations used
throughout this work.
\begin{lemma}\label{le0}
Let $P:=p_2(x, -i\nabla )$ be a linear operator on $L^2(\mathbb R^d)$ over the field $\mathbb C$,
where $p_2$ is a complex valued, quadratic
polynomial and specify its domain by
\begin{equation}
 \label{DP}
\mathcal D(P):= \{ f: \rea f, \im f\in C_0^\infty(\mathbb R^d)\}.
\end{equation}
Then $\overline P$ is the maximal extension of $P$ in the sense that
\begin{equation}
\mathcal D(\overline P)= \{ f\in L^2(\mathbb R^d): the\ distribution\ Pf\in L^2(\mathbb R^d)\}.
\end{equation}
\end{lemma}
\begin{proof} (sketch)
We define a mollifying delta sequence by
\begin{equation}
\varphi _n(x):=n^d\varphi (nx), \quad x\in\mathbb R^d, n\in\mathbb N,
\end{equation}
with $\varphi \in C_0^\infty$ and
$\varphi \geq 0,\
\varphi(x)=\varphi(-x),\ \int_{\mathbb R^d} \varphi (x)dx=1, \
\supp\varphi \subset \{|x|<1\}.$
Also, a sequence of radially symmetric cutoff function is defined by
\begin{equation}
\chi _n(x):=\chi \left(\frac{|x|}{n}\right),\quad x\in\mathbb R^d, n\in\mathbb N,
\end{equation}
with
$  \chi_n  \in C_0^\infty, \ 0\leq  \chi  \leq  1,\
   \supp \chi  \subset [0,1],
  \ \chi\big|_{[0,\frac12]}\equiv 1.$
\newpar
For $f\in L^2(\mathbb R^d)$ we define an approximating sequence in 
$\mathcal D(P)$ by
\begin{equation}
f_n(x):=\chi_n(x)(f\ast\varphi_n)(x), \quad n\in \mathbb N.
\end{equation}
We have to prove that for all  $f\in L^2(\mathbb R^d)$, with $Pf\in L^2(\mathbb
R^d)$, $f_{n}\rightarrow f$ in the graph norm ${\|f\|}_P:={\|f\|}_2
+{\|Pf\|}_2$.
We clearly have
\begin{equation}\label{conv1}
f_{n}\stackrel{n\rightarrow\infty}{\longrightarrow}f\;
\mbox{in }L^2(\mathbb R^d),
\end{equation}
and it remains to prove $Pf_n\rightarrow Pf$ in $L^2(\mathbb R^d)$. This
is now analogous to the proof of lemma 2.2 in \cite{ACD}, when extended to complex valued functions $f$.
A similar strategy is used again in the proof of lemma \ref{le} below.
\end{proof}

\begin{remark} Lemma \ref{le0} asserts that the \emph{minimal} and
\emph{maximal operators} defined by the expression
$P=p_2(x,-i\nabla)$ coincide. This fact is closely related to the
essential self-adjointness of Schr\"odinger operators. 
The lemma provides an elementary proof of
the well known fact that the Hamiltonian $H=-\Delta-|x|^2$ is
essentially self-adjoint on $C_0^\infty(\mathbb R^d)$, cf. 
corollary to theorem X.38 in \cite{ReSi2}; -- just
apply the lemma to $H$ with $\mathcal D(H)=C_0^\infty(\mathbb
R^d)$ and to $H^*\Big|_{\mathcal D(H)}$. On the other hand, it is
well known that $H=-\Delta+x^2-x^4$ is {\it not} essentially
self-adjoint on $C_0^\infty(\mathbb R)$, cf. example 1 of X.5 in
\cite{ReSi2}. Therefore, lemma \ref{le0} can, in general, {\it
not} be extended to higher order polynomials $p(x,-i\nabla)$. 
\end{remark}
With the above lemma we can now prove that the main technical assumption on the operator $Y$
(imposed in \cite{Da1}, \cite{ChFa}) is fulfilled.
\begin{proposition} \label{prop1}
Let $V_1=0$ and let the operator $Y$ be defined on
\begin{equation}
\mathcal D(Y):= \{ f\in L^2(\mathbb R^d): \Delta f, |x|^2f\in
L^2(\mathbb R^d)\}.
\end{equation}
\emph{\textbf{(a)}} Then its closure $\overline Y$ is the
infinitesimal generator of a $C_0$ - contraction semigroup on
$L^2(\mathbb R^d)$.
\newpar
\emph{\textbf{(b)}} Further, the operators $L_j$, $L_j^* :
\mathcal D(\overline Y) \rightarrow L^2(\mathbb R^d)$ satisfy
\begin{equation}
\label{as}\left<Yf,g\right> + \left<f,Yg\right>+
\sum_{j=1}^m\left<L_jf,L_jg\right>=0,\quad \forall f,g\in
\mathcal D(\overline Y),
\end{equation}
where $\left<\cdot,\cdot\right>$ denotes the standard scalar product on $L^2(\mathbb R^d)$.
\end{proposition}
\begin{proof}
First note that for $f\in \mathcal D(Y)$ the term $x\cdot \nabla
f$, which appears in $Yf$, is also in $L^2(\mathbb R^d)$. This can
be obtained by an interpolation argument. Further, $\mathcal D(Y)$
is dense in $L^2(\mathbb R^d)$, since $C_0^ \infty(\mathbb R^d)$
is. By Lemma \ref{le0} we have
$$
\mathcal D(\overline Y)= \{ f\in L^2(\mathbb R^d): Yf\in L^2(\mathbb R^d)\}.
$$
\newpar
\emph{Part (a):} The proof proceeds in several steps:
\newpar
\emph{Step 1:} We study the dissipativity of $Y$, which in our case is defined by
$$
\rea \left<Yf,f\right>\leq 0, \quad \forall f\in \mathcal D(Y).
$$
Since $H$ from (\ref{y}) is symmetric we obtain
$$
\rea \left<iHf,f\right>= 0, \quad \forall f\in \mathcal D(Y).
$$
Also we get
$$
- \rea \left<L_j^*L_jf,f\right> = - \left<L_jf,L_jf\right>\leq 0, \quad \forall f\in \mathcal D(Y).
$$
Thus $Y$ is dissipative and by theorem 1.4.5b of \cite{Pa} also its closure $\overline Y$ is.
\newpar
\emph{Step 2:}
Its adjoint is $Y^*=iH-\frac{1}{2}L$, with domain of definition $\mathcal D(Y^*)$.
We have $\mathcal D(Y^*)\supseteq \mathcal D(Y)$, since
$$
\left<Yf,g\right>=\left<f,Y^*g\right>, \quad \forall f, g\in \mathcal D(Y).
$$
As in step 1 we conclude that $Y^*\Big|_{\mathcal D(Y)}$ is
dissipative. We can now apply lemma \ref{le0} to
$P=Y^*\Big|_{\mathcal D(P)}$ with $\mathcal D(P)$ defined in
(\ref{DP}). Then $P$ is dissipative on $\mathcal D(P)\subseteq
\mathcal D(Y) \subseteq \mathcal D(Y^*)$. Since $Y^*$ is closed,
we have $\mathcal D(Y^*)=\mathcal D(\overline P)$, the domain of
the maximal extension. Thus $Y^*$ is dissipative on all of
$D(Y^*)$.
\newpar
\emph{Step 3:} Application of the Lumer-Phillips theorem (corollary 1.4.4 in \cite{Pa})
to $\overline Y$ (with ${(\overline Y)}^*=Y^*$) implies the assertion.
\newpar
\emph{Part (b):} We need to show: If $f$, $Yf\in L^2(\mathbb R^d)$, then
$L_jf$, $L^*_j f\in L^2(\mathbb R^d)$ follows. This can be easily seen from the fact that
$$
\frac12 \sum_{j} \left<L_jf,L_jf\right> =  - \rea \left< Yf,f \right> <\infty .
$$
Equation (\ref{as}) is then obtained by a simple computation.
\end{proof}
With these properties of $\overline Y$ (as stated in proposition
\ref{prop1}), theorem 3.1 of \cite{Da1} asserts that (\ref{lqds})
has a so called \emph{minimal solution}:
\begin{proposition} \label{prop2} \emph{[Davies '77]} There exists
a positive $C_0$ - semigroup of contractions $\Phi_t$ on
$\mathcal J_1^s$. Its infinitesimal generator is the
evolution operator $\mathcal L$, defined on a sufficiently large
domain $\mathcal D(\mathcal L)$, such that $\mathcal J_1^s
\supseteq \mathcal D(\mathcal L) \supseteq \mathcal D(Z)$.\\
Here, $Z: \mathcal D(Z)\rightarrow \mathcal J_1^s$ is the
maximally extended operator with domain
\begin{equation}
\mathcal D(Z)=\{\rho\in \mathcal J_1^s(\L^2 (\mathbb R^d)):
Z(\rho):=Y\rho + \rho Y^* \in \mathcal J_1^s(L^2(\mathbb
R^d))\}.
\end{equation}
\end{proposition}
{}From the above proposition we learn that the formal generator $\mathcal L$, in general, does not
unambiguously define a solution of the corresponding master
equation, in the sense of semigroups. Also, it is well known, that
the obtained minimal solution need not be trace preserving
(for nonconservative examples see \eg \ \cite {Da1,Ho}).\\
On the other hand, if the semigroup corresponding to the minimal
solution preserves the trace, it is the unique conservative QDS
associated to the abstract evolution problem (\ref{lqds}), \cf
\cite{CGQ, ChFa, FaRe, Ho}. 
We are going to prove now that in our case the minimal solution is indeed the unique QDS.
To this end, we need to introduce some more notation:
\newpar
{}From now on
we denote by
\begin{align*}
(M(g)f)(x):=g(x)f(x),\quad (C(g)f)(x):=(g \ast f)(x),\quad g\in
C_0^\infty(\mathbb R^d),
\end{align*}
a family of multiplication and convolution operators on
$L^2(\mathbb R^d)$, where $``\ast "$ is the usual convolution
w.r.t.~$x$. Further we define, for $n \in \mathbb N$, a family of
sets $\mathcal D_n\subset \mathcal J_1^s(L^2(\mathbb R^d))$ by
\begin{equation}
  \label{DDn}
  \mathcal D_n := \{ \sigma_n \in \mathcal J_1^s:\exists \rho\in \mathcal J_1^s \ s.t.\
  \sigma_n= M( \chi _n)C(\varphi _n) \rho \ C(\varphi_n)M(\chi _n) \},
\end{equation}
where $\chi_n$, $\varphi_n$ are the cutoff resp. mollifying functions defined in the proof of lemma \ref{le0} above.
For an operator $\rho \geq 0$ with kernel (\ref{fek}), the operator $\sigma_n$ has an integral kernel given by
\begin{align}
\sigma _n(x,y)& =  \ \chi _n(x)\varphi _n(x) \astx\rho (x,y)\asty \varphi _n(y)\chi _n(y)\nonumber\\
\label{ker}
& = \ \sum_{j\in\mathbb N} \lambda _j \ \varphi _{j, n}(x) \ \overline{\varphi _{j,n}(y)},
\end{align}
where $\varphi _{j, n}(x):=\chi _n(x)(\varphi_n \ast \psi _j)(x)
\in C_0^\infty(\mathbb R^d)$ and ${\|\varphi_{j,n}\|}_2 \le
{\|\psi_j\|}_2 = 1 $. Since $\sigma_n\ge0$ we get
\begin{equation}\label{est1}
 { \3norm\sigma_n\3norm}_1=\tr\sigma_n=\sum_{j\in \mathbb N}\lambda_j {\|\varphi_{j,n}\|}^2_2
  \le \sum_{j\in \mathbb N}\lambda_j={\3norm\rho\3norm}_1.
\end{equation}
The union of all sets $\mathcal D_n$ will be denoted by
\begin{equation}
\label{ds} \mathcal D_\infty := \bigcup_{n\in \mathbb N} \mathcal D_n.
\end{equation}
Also we shall write for the graph norm corresponding to $\mathcal L$
\begin{equation}
{\|\rho \|}_{\mathcal L}:= {\3norm \rho\3norm}_1+ {\3norm \mathcal L(\rho)\3norm}_1.
\end{equation}
Then the following technical result, which is a key point in the existence and uniqueness analysis, holds.
 \begin{lemma} \label{le} Let $V_1=0$. Then:\\
\emph{\textbf{(a)}} The set $\mathcal D_\infty$ is dense in $\mathcal J_1^s$.\\
\emph{\textbf{(b)}} $\mathcal D_\infty \subset \mathcal D(Z) \subset \mathcal D(\mathcal L)$.\\
\emph{\textbf{(c)}} The operator $\overline{{\mathcal
L\mid}_{\mathcal D_\infty}}$ is the maximal extension of $\mathcal
L$, in the sense that for each $\rho \in \mathcal J_1^s$,
with $\mathcal L(\rho)\in\mathcal J_1^s$, there exists a
sequence $\{\sigma_ n\}_{n\in\mathbb N}\subset \mathcal D_\infty$,
such that
\begin{equation}
\lim_{n\rightarrow \infty }{\|\rho -\sigma _n\|}_{\mathcal L} = 0.
\end{equation}
\end{lemma}
\begin{proof} The proof is deferred to the appendix.
\end{proof}
\newpar
\begin{remark} For all $\rho \in \mathcal J_1^s$, $\mathcal
L(\rho)$ can be defined (at least) as an operator $\mathcal
L(\rho): C_0^ \infty(\mathbb R^d) \rightarrow \mathcal D'(\mathbb
R^d)$, the space of distributions. For $\mathcal
L(\rho)\in\mathcal J_1^s$ to hold, first of all an
appropriate extension has to exist, such that $\mathcal L(\rho)
\in \mathcal B (L^2(\mathbb R^d))$.
\end{remark}
We are now in the position to state our first main theorem:
\begin{theorem} \label{lth} Let $V_1=0$.
The evolution operator $\mathcal L$ generates on $\mathcal J_1^s$ 
a conservative quantum dynamical semigroup of contractions
$\Phi_t(\rho)=e^{\mathcal L t}\rho$. This QDS yields the unique
mild solution, in the sense of semigroups, for the abstract
evolution problem (\ref{lqds}).
\end{theorem}
\begin{proof}
Existence of $\Phi_t(\rho)=e^{\mathcal L t}\rho$ is guaranteed by
proposition \ref{prop2}. As a semigroup generator $\mathcal L$ is
closed, and by lemma \ref{le} it is the maximally extended
evolution operator. This implies uniqueness of the semigroup.
Complete positivity then follows from Stinespring's theorem
\cite{Sti, AlFa}.
\smallskip \\
It remains to prove the conservativity for the obtained QDS. This will be done by using a similar argument as in
the proof of theorem 3.2 in \cite{Da1}:
\newpar
\emph{Step 1:} For the special case $\rho_0\in \mathcal D(\mathcal
L)$ the trajectory $\Phi_t(\rho_0)$ is a classical solution (in
the sense of semigroups, \cf \cite {Pa}), \ie $\Phi_t(\rho_0)\in
C^1([0,\infty), \mathcal J_1( L^2(\mathbb R^d)))$ and $\Phi_t(\rho_0)\in \mathcal
D(\mathcal L)$, $\forall$ $t\geq 0$. Hence $\tr \Phi_t(\rho_0) \in
C^1([0,\infty), \mathbb R)$ and we calculate for $t\geq 0$:
\begin{align}
\label {con} \frac{d}{dt}\tr \Phi_t(\rho_0)= \tr \frac{d}{dt} \Phi_t(\rho_0) =
\tr \mathcal L(\Phi_t(\rho_0))=0.
\end{align}
To justify the last equality we note that $\mathcal D_\infty$ is $\| \cdot \|_{\mathcal L}$ - dense in
$\mathcal D(\mathcal L)$, by lemma \ref{le} (c). Thus we can approximate $\Phi_t(\rho_0)$, for
every fixed $t\geq0$, by an appropriate sequence $\{ \sigma_n \} \subseteq \mathcal D_\infty $.
Since $\mathcal D_\infty$ is included in the domain of each ``term" (\ref{terms}) of the operator $\mathcal L$
(as the proof of lemma \ref {le} (b) shows), the cyclicity of the trace yields
$\tr \mathcal L(\Phi_t(\rho_0))=0$. Equation (\ref{con}) then implies
$$
\tr \Phi_t(\rho_0)=\tr \rho_0=0, \quad \forall \rho_0\in \mathcal D(\mathcal L), t\geq0.
$$
\newpar
\emph{Step 2:} The general case $\rho_0\in \mathcal J_1^s( L^2(\mathbb R^d))$
(\ie $\Phi_t(\rho_0)$ is a mild solution) follows
{}from step 1 and the fact that $\mathcal D(\mathcal L)$ is dense in
$\mathcal J_1^s( L^2(\mathbb R^d))$.
\end{proof}

{}From the above theorem, we obtain the the following corollary:

\begin{corollary} \label{cor0}
For $\rho \in \mathcal D(\mathcal L)$ let
\begin{equation}
\tilde{\mathcal L}(\rho):=\mathcal L(\rho)+\mathcal L_p(\rho),
\end{equation}
where
\begin{equation}
\mathcal L_p(\rho):= - i[V_1, \rho] +
\sum_{j=m+1}^{\infty} L_j \rho L_j^*-\frac{1}{2}\left[L_j^*L_j,
\rho \right]_+,
\end{equation}
with $V_1\in L^\infty (\mathbb R^d)$, $L_j\in \mathcal
B(L^2(\mathbb R^d))$ and the sum converges in $\mathcal
B(\mathcal J_1^s( L^2(\mathbb R^d)))$. Then the perturbed operator
$\tilde{\mathcal L}$ again uniquely defines a conservative QDS of contractions.
\end{corollary}
\begin{proof} Existence and uniqueness of the $C_0$-semigroup follows from standard
perturbation results, \cf \cite{Pa}. To prove conservativity of
the perturbed QDS, let $\rho(t)$ denote the solution of
$$
\frac{d}{dt}\rho=\tilde{\mathcal L}(\rho),\quad \rho(0)=\rho_0.
$$
The conservativity then follows from Duhamel's representation
\begin{align}\label{duh}
\rho(t)=\Phi_t(\rho_0)+\int_0^t\Phi_{t-s}(\mathcal L_p(\rho(s))) \ ds,
\end{align}
by noting that $\tr (\mathcal L_p(\rho))=0$. All other properties can be established
by the same procedure as in theorem 1 of \cite{AlMe} or by a Picard iteration.
\end{proof}
\begin{remark}
An alternative approach to prove theorem \ref{lth} could be to verify the sufficient conditions of
\cite{ChFa}. In fact their assumptions A1 and A2 are simple consequences of our lemma \ref{le0} and
proposition \ref{prop1}. For their third condition A3 however, one would need to prove that
$C_0^\infty(\mathbb R^d)$ is a core for $Y^2$, defined on
\begin{equation}
\mathcal D(Y^2):=\{ f\in \mathcal D(\overline{Y}): \overline{Y} f\in \mathcal D(\overline{Y})\}.
\end{equation}
With considerable more effort, the proof should be possible by extending the strategy of lemma \ref{le0}.
However, one can expect quite cumbersome calculations.
\end{remark}


\section {Local-in-time existence of the mean field QDS} \label{nlinl}

We shall now prove existence and uniqueness of local-in-time
solutions for the nonlinear evolution problem
\begin{equation} \label{nlqds}
\left \{
\begin{split}
& \frac{d}{dt} \rho = \mathcal L(\rho), \quad t>0\\
& \rho(0)=\rho_0\in \mathcal J_1^s.
\end{split}
\right.
\end{equation}
Here, the nonlinear map $\mathcal L$ is given by
\begin{equation}
\mathcal L(\rho) := -i\left[-\frac{\Delta}{2} + V[\rho]- i \mu {[x, \nabla]}_+, \rho
\right] + A(\rho),
\end{equation}
where the self-consistent potential $V[\rho]$ is given as in (\ref{pot}) and $A(\rho)$ is
the Lindblad operator defined by (\ref{lind}) and (\ref{lind1}).
\newpar
To this end, we shall prove that the linear evolution problem
(\ref{lqds}) not only defines a $C_0$-semigroup in
$\mathcal J_1^s$ (guaranteed by theorem \ref{lth}) but
also in an appropriate energy space. This is a parallel procedure
(apart from severe technical difficulties) to solving the
\emph{Schr\"odinger-Poisson} equation in $H^1(\mathbb R^d)$, \cf \cite{GiVe}.\\
Note that Davies' construction of a minimal QDS is valid
\emph{only} in $\mathcal J_1$. Hence, the required additional
regularity of $\Phi_t(\rho_0)$ has to be established explicitly.
Also, one has to prove separately that this nonlinear model conserves the positivity and the trace of $\rho$.
\newpar
In the following, we shall restrict ourselves to
the physical most important case of $d=3$ spatial dimensions.
\newpar
Let us start by introducing the following definitions:
\begin{definition} The \emph{kinetic energy} of a density matrix operator $\rho \in
\mathcal J_1^s$ is defined by
\begin{equation}
E^ {kin}[\rho]:=
\frac{1}{2}\tr(\sqrt{-\Delta}\,\rho\sqrt{-\Delta}),
\end{equation}
where $\sqrt{-\Delta}$ denotes a \emph{pseudo-differential operator} with symbol $|\xi|$,
$\xi \in \mathbb R^d$, \ie
\begin{equation}
\sqrt{-\Delta} f(x) := \frac{1}{(2\pi)^d} \int_{\mathbb R^d} |\xi| (\mathcal F f)(\xi)
e^ {i \xi\cdot x} d\xi, \quad \forall f\in H^1(\mathbb R^d).
\end{equation}
Further, we define the \emph{external} and the \emph{self-consistent potential energy}
of $\rho \in \mathcal J_1^s$ by
\begin{equation}
E^ {ext}[\rho]:= \frac{1}{2}\tr ( |x| \, \rho \, |x|) ,\quad 
E^ {sc}[\rho]:= \frac{1}{2}\tr(\phi[\rho]\rho).
\end{equation}
The total energy will be denoted by
\begin{equation}
E^ {tot}[\rho]:= E^ {kin}[\rho]+E^ {ext}[\rho]+E^ {sc}[\rho].
\end{equation}
In the sequel we shall work in the following energy space $\mathcal E$:
\begin{equation}
\mathcal E:= \{ \rho \in \mathcal J_1^s:  \sqrt{1-\Delta+|x|^2}\,\rho\sqrt{1-\Delta+|x|^2}\in 
\mathcal J_1^s \},
\end{equation}
equipped with the norm
\begin{align}
{\|\rho \|}_{\mathcal E}:= {\3norm\, \sqrt{1-\Delta+|x|^2}\,\rho\sqrt{1-\Delta+|x|^2} \, \3norm}_1
\end{align}
\end{definition}
This energy norm is a generalization of the one defined in \cite{BDF}. 
In case $\rho$ is indeed a \emph{physical} state,  \ie $\rho\geq 0$, and if in addition 
$\rho\in \mathcal D_\infty$, one easily gets 
\begin{equation}
{\|\rho \|}_{\mathcal E} = 
{\3norm \rho \3norm}_1+ {\3norm\sqrt{-\Delta}\,\rho \sqrt{-\Delta} \, \3norm}_1 + {\3norm\, |x| \,
\rho \, |x| \, \3norm}_1, \quad \forall \rho\in \mathcal D_\infty,\, \rho\geq 0.
\end{equation}
Hence, a density argument, similar to lemma~\ref{le} (c), implies for all
$\rho\geq 0$ that $\rho\in \mathcal E$ is \emph{equivalent} to
$\rho\in \mathcal J_1^s$ and  $E^ {kin}[\rho]+E^ {ext}[\rho]<\infty$.\\ 
We further remark that in the above definitions we neglected the term $-i\mu
[x,\nabla ]_+$,  which appears in the generalized (or adjusted) Hamiltonian
operator (\ref{ham}) of our system. Thus, even in the linear case, we have
$E^{tot}[\rho ]\not = \tr(H\rho )$. The latter term would be the more common
definition for the energy of the system. We note that we shall use
$E^{tot}[\rho]$ only for deriving a-priori estimates and towards this end
$E^{tot}[\rho]$ is the more convenient expression.
\newpar
\begin{remark} Using the cyclicity of the trace, one formally obtains
the more common expression for the kinetic energy of a physical state $\rho\geq 0$:
\begin{equation}
E^{kin}[\rho]:= \frac{1}{2}\tr(\sqrt{-\Delta}\,\rho\sqrt{-\Delta})=  \frac{1}{2}\tr(-\Delta\rho)\geq 0.
\end{equation}
However, these two expressions for $E^{kin}[\rho]$ are not fully
equivalent, since $\Delta\rho\in \mathcal J_1^s$ requires
more regularity on $\rho$ than just requiring $\sqrt{-\Delta}\rho
\sqrt{-\Delta}\in \mathcal J_1^s$. (For more details see
\eg \ \cite{Ar} and the references given therein.) We further
remark that if the kernel of $\rho$ is given as in (\ref{fek}) the
kinetic energy reads
\begin{equation}
E^{kin}[\rho]=\frac{1}{2}\sum_{j\in\mathbb N} \lambda _j \, {\|\, \nabla \psi_j\, \|}_2^2\geq 0.
\end{equation}
Similarly we get that for physical states $\rho \geq 0$ it holds $E^{ext}[\rho]\geq 0$, as well as 
$E^ {sc}[\rho]\geq0$, since $\rho\geq 0$ implies $n[\rho]\geq0$ and hence $\phi[\rho]\geq 0$, by (\ref{spot}). 
Finally, note the additional factor $1/2$ in front of the term $E^{sc}[\rho ]$,
which does not appear in the Hamiltonian (\ref{ham}), (\ref{pot}).
It is due to the self-consistent nonlinearity, \cf \cite{Ar}. 
\end{remark}
\newpar
Using these definitions, we will now prove that the sum of kinetic and (external) potential energy
is continuous in time during the linear evolution.
\begin{lemma} \label{elem}
Let $V_1=0$ and $\rho_0\in \mathcal E$, then
\begin{equation}
(E^ {kin}+ E^ {ext})[\rho(t)]\in C([0,\infty);\mathbb R),
\end{equation}
where $\rho(t):= \Phi_t(\rho_0)\in C([0,\infty), \mathcal J_1^s)$ 
denotes the unique QDS for the linear evolution problem,
given by (\ref{lqds}).
\end{lemma}
\begin{proof}
First, we note that each $\rho \in \mathcal E\subset\mathcal J_1^s$ can be uniquely decomposed
into: $\rho =\rho_1-\rho_2$, where
\begin{equation}\label{edec}
\rho_{1,2}:= \Lambda^{-1}(\Lambda \rho \Lambda)^{\pm}\Lambda^{-1},\quad \Lambda:= \sqrt{1-\Delta+|x|^2},
\end{equation}
and $(\Lambda \rho \Lambda)^{\pm}$ denotes the positive resp. negative part of 
$(\Lambda \rho \Lambda)\in \mathcal J_1^s$. It holds: $\rho_{1,2} \geq 0$, as well as 
$\rho_{1,2}\in \mathcal E$. Using this decomposition for the intial data $\rho_0\in \mathcal E$ 
and since $\Phi_t$ preserves positivity, we can restrict ourselves in the following to the
case $\rho_0\geq 0$, hence $\rho(t)\geq 0$. 
The idea is now to derive a differential inequality for 
$E^{kin}+E^{ext}$ from (\ref{lqds}).
\newpar
Let us define some energy functionals for positive $\rho \in \mathcal J_1^s$:
\begin{align}\label{enfu}
E^ {kin}_{k,l}[\rho]:= -\frac{1}{2}\tr(\partial_k \rho \partial_l),  \quad
E^ {ext}_{k,l}[\rho]:= \frac{1}{2}\tr(x_k \rho x_l),
\end{align}
with $k,l=1,\dots ,d$. For $\rho\in\mathcal D_\infty$, the cyclicity of the trace implies
\begin{align}
\label{enre}
E^ {kin}[\rho]=\sum_{k=1}^d E^{kin}_{k,k}[\rho],\quad
E^ {ext}[\rho]=\sum_{k=1}^d E^{ext}_{k,k}[\rho]
\end{align}
and, by a density argument, the formulas (\ref{enre}) also hold
for $\rho\in\mathcal E$.
\newpar
\emph{Step 1:} We apply the operators $x_k$, $\partial_k$ (from
left and right) to (\ref{lqds}) and take traces. A lengthy but
straightforward calculation, using the cyclicity of the trace and
setting w.r.o.g. $\tr \rho(t)=1$, yields for the kinetic energy:
\begin{align} \label{dte1}
\sum_{k=1}^d \frac{d}{dt}E^ {kin}_{k,k}= & \
\frac{1}{2}\sum_{k=1}^d\sum_{j=1}^m |\alpha_{j,k}|^2-4\mu \sum_{k=1}^d E^ {kin}_{k,k}-
2  \sum_{k,l=1}^d \sum_{j=1}^m \rea(\alpha_{j,k}\overline{\beta_{j,l}})E^ {kin}_{k,l}\nonumber\\
& -  \sum_{k,l=1}^d  i\sum_{j=1}^m \im(\alpha_{j,k}\overline{\alpha_{j,l}})\tr(\partial _k\rho x_l )+
\im(\alpha_{j,k}\overline{\gamma_{j}})\tr(\rho \partial _k)\nonumber\\
& +i\left(\frac{d}{2}+\sum_{k=1}^d\tr(\partial_k\rho x_k)\right).
\end{align}
For the external energy we obtain:
\begin{align}\label{dte2}
\sum_{k=1}^d \frac{d}{dt}E^ {ext}_{k,k}= & -\frac{1}{2}\sum_{k=1}^d\sum_{j=1}^m |\beta_{j,k}|^2+
4\mu \sum_{k=1}^d E^ {ext}_{k,k}+
2\sum_{k,l=1}^d \sum_{j=1}^m  \rea(\alpha_{j,k}\overline{\beta_{j,l}})E^ {ext}_{k,l}\nonumber\\
& + i \sum_{k,l=1}^d \sum_{j=1}^m  \im(\beta_{j,k}\overline{\beta_{j,l}})\tr(\partial _k\rho x_l )+
\im(\beta_{j,k}\overline{\gamma_{j}})\tr(\rho x_k)\nonumber\\
& -i \left(\frac{d}{2}+\sum_{k=1}^d\tr(\partial_k\rho x_k)\right).
\end{align}
\emph{Step 2:} These equations are not closed in $E^ {kin}$ and $E^ {ext}$.
To circumvent this problem, we shall use interpolation arguments: First, note that
$(\partial_k\rho \partial_k)\in \mathcal J_1$, iff $(\partial_k \sqrt{\rho}) 
\in \mathcal J_2$, \cf \cite{ReSi1}.
Thus we can estimate
$$
{\3norm \rho \partial_k \3norm}_1^2 \leq \ {\3norm \sqrt{\rho} \3norm}_2^2 \ 
{\3norm \sqrt{\rho} \partial_k \3norm}_2^2 \
= \  {\3norm \rho \3norm}_1 \ {\3norm \partial_k \rho \partial_k \3norm}_1.
$$
Likewise, we get
$$
{\3norm  \partial_k \rho x_l \3norm}_1^2 \leq \  {\3norm \partial_k \sqrt{\rho} \3norm}_2^2 \ 
{\3norm \sqrt{\rho} x_l \3norm}_2^2 \
= \  {\3norm \partial_k \rho \partial _k \3norm}_1 \ {\3norm x_l \rho x_l \3norm}_1
$$
and one easily derives analogous estimates for the off-diagonal energy-terms $E^{ext/kin}_{k,l}$.
Hence, estimating term-by-term in (\ref{dte1}), (\ref{dte2}), we finally obtain
$$
 \left| \frac{d}{dt}\sum_{k=1}^d (E^ {kin}_{k,k}+E^ {ext}_{k,k})[\rho(t)]
 \right| \leq \  K\sum_{k=1}^d (E^ {kin}_{k,k}+E^ {ext}_{k,k})[\rho(t)],
$$
with some generic constant $K\geq 0$. Applying Gronwall's lemma then gives the desired result.
\end{proof}
This lemma directly leads to our next proposition:
\begin{proposition} \label{prop3}
Assume that $\rho_0\in \mathcal E$ and $V_1\in L^\infty(\mathbb R^d)$ s.t. additionally
$\nabla V_1\in L^q(\mathbb R^d)$, for some $3\leq q\leq \infty$. Then
\begin{equation}
\Phi_t(\rho_0)\in C([0,\infty), \mathcal E), 
\end{equation}
where $\Phi_t(\rho_0)$ denotes the unique linear QDS corresponding
to (\ref{lqds}).
\end{proposition}
\begin{proof} The proof is based on a generalization of Gr\"umm's theorem.
As described in the proof of lemma \ref{elem} above, we only need to consider, w.r.o.g., 
the case $\rho(t)\geq0$.
\newpar
\emph{Step 1:} At first, one proves that for all
$f$, $g\in L^2(\mathbb R^d)$ and $s \ge 0$,
\begin{equation}\label{weq}
\lim_{t\rightarrow s}\langle f,\Lambda\rho(t)\Lambda g\rangle =
 \langle f,\Lambda\rho(s) \Lambda g \rangle ,
\end{equation}
where $\left<\cdot,\cdot\right>$ denotes the standard $L^2(\mathbb R^d)$
scalar product. Choosing two sequences $\{f_n\}$, $\{g_n\} \subset C_0^\infty
(\mathbb R^d)$, 
s.t.
$f_n\stackrel{n \rightarrow \infty}{\longrightarrow} f$,
$g_n\stackrel{n \rightarrow \infty}{\longrightarrow} g$ in
$L^2(\mathbb R^d)$ the assertion then 
follows from a fairly standard approximation procedure.
\newpar
\emph{Step 2:} Let $V_1=0$ first. By theorem 2.20 in \cite{Si} (a generalization of Gr\"umm's theorem),
step 1 and the continuity of
$$
{\3norm\rho(t)\3norm}_1+2(E^{kin}+E^{ext})[\rho(t)]= {\3norm\Lambda\rho(t) \Lambda\3norm}_1
$$
(cf. lemma \ref{elem}) imply
$$
\lim_{t\rightarrow s}{\3norm\Lambda\:\left( \rho(t)-\rho(s)\right) \Lambda \3norm}_1 
= 0, \quad \forall s \ge 0.
$$
Thus
$$
\Lambda\rho(t)\Lambda \in C([0,\infty), \mathcal J_1^s(L^2(\mathbb R^d)))
$$
and the proposition is proved. The case $V_1\not=0$ can now be included by
a standard perturbation result, \cf \cite{Pa} under the additional assumption that
$\nabla V_1\in L^q(\mathbb R^d)$, for some $3\leq q \leq \infty$,
\cf \cite{Ar} for the detailed calculations.
\end{proof}
As a remaining preparatory step, the following lemma states an
important property of the nonlinear mean field
potential $\phi[\rho]$.
\begin{lemma} \label{lip}
Let $\rho\in \mathcal E$ and $d=3$, then $ \phi [\rho ]\in L^\infty (\mathbb R^3)$. Moreover,
the operator $[\phi[\rho], \rho]$ is a local Lipschitz map from $\mathcal E$ into itself.
\end{lemma}
\begin{proof} Once again we decompose $\rho = \rho_1 -\rho_2$ s.t. $\rho_{1,2}\geq 0$ and 
$\rho_{1,2} \in \mathcal E$, as given in (\ref{edec}). In $d=3$, we explicitly get from (\ref{spot})
$$
\phi [\rho_{j}]  = - \frac{1}{4\pi  |x|}\ast n[\rho_{j} ],\quad \nabla \phi [\rho_{j}]  =  
\frac{x}{4\pi  |x|^3}\ast n[\rho_{j} ] ,\quad j=1,2.
$$
Therefore, the Hardy-Littlewood-Sobolev inequality and the generalized Young inequality, \cf \cite{ReSi2}, imply 
for $j=1,2$:
\begin{align*}
\phi [\rho_{j}]\in L^3_w (\mathbb R^3) \cap L^p(\mathbb R^3),\quad 3<p<\infty, 
\end{align*}
as well as
\begin{align*}
\nabla \phi [\rho_{j}]\in L^{3/2}_w (\mathbb R^3) \cap L^p(\mathbb R^3),\quad 3/2<p<\infty.
\end{align*}

Here, $L^{p}_w $ denotes the weak $L^p$-spaces, \cf \cite{ReSi2}. Hence, by a
Sobolev imbedding, we obtain $\phi[\rho]\in L^\infty(\mathbb R^d)$. Similar
arguments as given in the proof of lemma 3.11 in \cite{Ar} then imply that
$[\phi[\rho],\rho]$ is a  local Lipshitz map in the energy space $\mathcal E$.
To this end we first estimate
$$
  \3norm\Lambda \phi[\rho] \rho \Lambda \3norm_1 \le
  \3norm\Lambda \phi[\rho] \Lambda^{-1} \3norm_\infty \;
  \3norm\Lambda \rho \Lambda \3norm_1
$$
and use the assumption $\Lambda\rho \Lambda\in \mathcal J_1$.
For the first factor on the r.h.s.~one calculates for $f \in C_0^\infty(\mathbb R^3)$:
$$
  \|\Lambda \phi[\rho] \Lambda^{-1} f\|^2_2=
  \|\nabla \left(\phi[\rho] \Lambda^{-1} f\right)\|^2_2+
  \|\sqrt{1+|x|^2} \phi[\rho] \Lambda^{-1} f\|^2_2
$$
We rewrite the operator of the first term on the r.h.s.~as
$$
\nabla \left(\phi[\rho] \Lambda^{-1}\right) = 
\left[((\nabla \phi[\rho])+\phi[\rho] \nabla) (1-\Delta)^{-1/2}\right]\;
\left[(1-\Delta)^{1/2}\Lambda^{-1}\right],
$$ 
where both factors are in $\mathcal B(L^2(\mathbb R^3))$. The first factor is
bounded since $\nabla \phi[\rho] \in L^3(\mathbb R^3)$ and since
$(1-\Delta)^{-1/2}$  is a bounded map from $L^2(\mathbb R^3)$ into $H^1(\mathbb
R^3)\hookrightarrow  L^6(\mathbb R^3)$, due to a Sobolev imbedding. 
\newpar
Summarizing we obtain
$$
{\| [\phi[\rho], \rho] \|}_{\mathcal E}\leq C {\| \rho \|}^2_{\mathcal E}, \quad \forall \rho \in \mathcal E,
$$
and the Lipshitz continuity then follows in a straightforward way.
\end{proof}
We remark that the nonlinear map $\rho \mapsto [\phi[\rho],\rho]$
is continuous in $\mathcal E$, but \emph{not} in $\mathcal J_1^s
(L^2(\mathbb R^3))$ and this is the reason why we need to work
in the energy space $\mathcal E$. However, the linear evolution
problem (\ref{lqds}) in general does \emph{not} generate a
\emph{contractive} QDS on $\mathcal E \subset \mathcal J_1$,
except in the case of a unitary dynamic (\ie $L_j=0$). Hence, in
order to obtain a global-in-time (nonlinear) existence and
uniqueness result, we can not apply the results of \cite{AlMe},
which would require contractivity of the linear QDS in $\mathcal E$. \\
In the nonlinear evolution problem (\ref{nlqds}) the situation is
even worse. Already in the case of a unitary time-evolution only
$E^{tot}[\rho(t)]$ is conserved (for $\mu=0$), whereas ${\|
\rho(t) \|}_{\mathcal E}$ is not, due to the possible energy
exchange between the potential and the kinetic parts. Hence a
unitary but self-consistent evolution problem does not generate a
contractive semigroup in $\mathcal E$ either.
\newpar
With the above results, we are able to state the following local-in-time result:
\begin{theorem}\label{thn1}
Let $\rho_0\in \mathcal E$, $d=3$ and $V_1\in L^\infty(\mathbb R^3)$ s.t.
$\nabla V_1\in L^q(\mathbb R^3)$, for some $3\leq q\leq \infty$, then:
\newpar
\emph{\textbf{(a)}} Locally in time, the nonlinear evolution
problem (\ref{nlqds}) has a unique mild solution
$\tilde{\Phi}_t(\rho_0)\in C([0,T), \mathcal E)$, where
$\tilde{\Phi}_t(\cdot)$ denotes the nonlinear semigroup obtained
by perturbing the linear QDS with the Hartree potential. This
self-consistent potential satisfies: $\phi\in C([0,T); C_b(\mathbb
R^3))$. The map $ \rho_0 \mapsto \tilde{\Phi}_t (\rho_0)$ is
Lipschitz continuous on some (small enough) ball
${\{\|\rho-\rho_0\|}_{\mathcal E} < \varepsilon\}\subset\mathcal
E$, uniformly for $0 \le t \le T_1 < T$. Further, if the maximum
time of existence $T>0$ is finite, we have
\begin{equation}\label{blow}
\lim_{t\nearrow T}{\| \tilde{\Phi}_t(\rho_0)\|}_{\mathcal E}=\infty.
\end{equation}
\emph{\textbf{(b)}} For $\mathcal L(\rho_0) \in \mathcal E$ we
obtain a classical solution $\tilde{\Phi}_t(\rho_0) \in
C^1([0,T),\mathcal
E)$. \\
\emph{\textbf{(c)}} The semigroup $\tilde{\Phi}_t$ is conservative.\\
\emph{\textbf{(d)}} $\,$The semigroup $\tilde{\Phi}_t$ is
positivity preserving and contractive on $\mathcal J_1^s
(L^2(\mathbb R^3))$. Hence, it furnishes a nonlinear QDS:
$\tilde{\Phi}_t: \mathcal E\rightarrow \mathcal E\subset
\mathcal J_1^s$.
\end{theorem}
\begin{proof}
\emph{Part (a, b):} By proposition \ref{prop3} the unique
conservative QDS $\Phi_t$, obtained from theorem \ref{lth}, also
maps the energy space $\mathcal E$ into itself. Lemma \ref{lip}
and a standard perturbation result (\cf theorem 6.1.4 in
\cite{Pa}) then yield the local-in-time existence of a solution
for the nonlinear, \ie mean field problem. The continuity of
$\phi$ follows from the proof of lemma \ref{lip}, using
$\tilde{\Phi}_t(\rho_0)\in C([0,T); \mathcal E)$.
The local Lipschitz continuity of the map $ \rho_0 \mapsto \tilde{\Phi}_t (\rho_0)$
follows from theorem 6.1.2 in \cite{Pa} and the uniform lower bound for the
existence time of trajectories $\tilde{\Phi}_t (\rho)$ that start in the neighborhood of
$\rho_0$ (cf. proof of theorem 6.1.4 in \cite{Pa}).
\newpar
\emph{Part (c):} The proof follows from Duhamel's representation,
analogous to (\ref{duh}).
\newpar
\emph{Part (d):} Having in mind the result of part (a), we
consider the nonlinear evolution problem (\ref{nlqds}) as a linear
evolution problem with time-dependent Hamiltonian and write it in
the following form:
\begin{equation}\label{tdh}
\left \{
\begin{split}
&\frac{d}{dt}\, \rho = -i \left[H, \rho\right] + A(\rho)-i[\phi(t), \rho], \quad t>0,\\
&\rho(0)=\rho_0\geq 0.
\end{split}
\right.
\end{equation}
Here, $\phi\in C([0,T); C_b(\mathbb R^3))$ is the self-consistent
potential $\phi[\rho]$. To prove the assertions of part (d), we
shall approximate $\phi(t)$ on $[0,T_1]$, $T_1<T$, by the
piecewise constant potential:
$$
\vartheta(t):=\phi(t_n),\quad t_n\leq t<t_{n+1},\ 0\leq n\leq N-1,
$$
with the uniform grid points: $t_n=n\Delta t$, $\Delta t = T_1/N$.
Hence, $\rho(t)$, $t\in [0,T_1]$ is approximated by $\varsigma_N\in
C([0,T_1]; \mathcal J_1^s(L^2(\mathbb R^3)))$, solving
\begin{equation}\label{appr}
\left \{
\begin{split}
&\frac{d}{dt} \, \varsigma_N = -i \left[H, \varsigma_N \right] +
A(\varsigma_N)-
i[\vartheta(t), \varsigma_N], \quad t>0,\\
&\varsigma_N(0)=\rho_0\geq 0.
\end{split}
\right.
\end{equation}
Since $\vartheta(t)\in C_b(\mathbb R^3)$, corollary \ref{cor0}
applies to the generator
in (\ref{appr}) on each time-intervall $[t_n, t_{n+1}]$. In summary we have the following facts:\\
$\phi$ is uniformly continuous on $[0,T_1]$ w.r.t. ${\|\cdot\|}_\infty$,
the solutions of (\ref{tdh}) satisfies:
${\3norm\rho(t) \3norm}_1\leq K$, on $0\leq t\leq T_1$,
and the propagator corresponding to (\ref{appr}) is contractive on
$\mathcal J_1^s(L^2(\mathbb R^3))$.
\newpar
With these ingredients it is standard to verify that
$$
\lim_{N\rightarrow \infty} \varsigma_N = \rho, \quad
\mbox{in\, $C([0,T_1];\mathcal J_1^s(L^2(\mathbb R^3)))$,}
$$
\cf the proof of theorem 1 in \cite{AlMe} \eg. Hence, the positivity of
$\rho(t)=\tilde{\Phi}_t(\rho_0)$ follows from the positivity of $\varsigma_N(t)$.\\
Analogously, the contractivity of the propagator corresponding to (\ref{appr}) implies the
contractivity of $\tilde{\Phi}_t(\rho_0)$ in $\mathcal J_1^s(L^2(\mathbb R^3))$.
\end{proof}
\begin{remark} If no confinement potential is present and
$\im(\alpha_{j,k}\overline{\alpha}_{j,l})=0$, $\forall j,k,l$, then theorem \ref{thn1}
also holds in the kinetic energy space $\mathcal E^{kin}$.
In particular, this is true for the QFP equation, where one can
derive an exact ODE for the kinetic energy, \cf \cite{ALMS}.
\end{remark}
In the next section we shall derive a-priori estimates on
$\tilde{\Phi}_t(\rho)$ to prove the global-in-time existence of a
conservative QDS for the mean field problem.


\section {A-priori estimates and global existence of the mean field QDS} \label{nling}

{}From theorem \ref{thn1}, we already know that ${\3norm
\rho(t)\3norm}_1= {\3norm \rho_0\3norm}_1$, for $0\leq t<T$. It remains to
prove an a-priori estimate on the energy of the nonlinear system.
As a preliminary step, we introduce a generalized version of the
\emph{Lieb-Thirring inequality}:
\begin{lemma} Assume $d=3$ and let $\rho \in \mathcal J_1^s$, $\rho\ge 0$
be s.t. $E^{kin}[\rho ]<\infty $.
Then the following estimate holds:
\begin{align}\label{lth1}
{\| n [\rho ] \|}_p \, \leq \, K_p \, {\3norm \rho \3norm}_1^{\theta} E^{kin}[\rho ] ^{1-\theta },
\quad  1\leq p\leq 3,
\end{align}
with
\begin{equation}
\theta := \frac{3-p}{2p}.
\end{equation}
\end{lemma}
\begin{proof} The proof is given in the appendix of \cite{Ar}, \cf also \cite{LiPa}.
\end{proof}
In the sequel this estimate will be used to derive an a-priori
bound for the total energy.
\begin{proposition} Assume $\rho_0\in \mathcal E,\, \rho_0 \ge 0$ and $d=3$.
Then there exists a $K>0$ such that
\begin{equation}
\label{noname2}
 E^{tot}[ \rho(t) ] \le e^{Kt}E^{tot}[\rho_0],
\quad 0\leq t<T,
\end{equation}
where $\rho(t) :=\tilde{\Phi}_t(\rho_0)$, denotes the unique
local-in-time solution of the nonlinear evolution problem
(\ref{nlqds}).
\end{proposition}
\begin{proof} Since $\tilde{\Phi}_t$ is positivity preserving, we assume w.r.o.g.
$\rho_0\geq 0$ and hence have $\rho(t)\geq 0$, for all $0\leq t<
T$. The idea is again to derive a differential inequality for
$E^{tot}$. We first consider a classical solution
$\tilde{\Phi}_t(\rho_0) \in C^1([0,T),\mathcal E)$ obtained from
an initial condition with $\mathcal L(\rho_0) \in \mathcal E$.
\newpar
\emph{Step 1:} We calculate the time derivative of the total energy, using the
short notation $\dot \rho \equiv \frac{d}{dt}\rho$:
\begin{align}\label{apr}
 \frac{d}{dt} E^{tot}[\rho]= & \ \frac{d}{dt} \tr\left(-\frac{1}{2}\sqrt{-\Delta} \rho \sqrt{-\Delta} 
  + \frac{1}{2}|x| \rho |x| + \phi [\rho]\rho \right)
- \frac{1}{2} \frac{d}{dt} \tr(\phi [\rho] \rho) \nonumber\\
= & \ \tr\left( -\frac{1}{2}\sqrt{-\Delta} \dot \rho \sqrt{-\Delta}    + \frac{1}{2}|x| \dot \rho |x|
+\phi [\rho] \dot \rho \right) + \tr (\dot \phi [\rho] \rho) \nonumber\\
& -  \frac{1}{2} \frac{d}{dt} \tr(\phi [\rho] \rho).
\end{align}
For our classical solution $\rho(t)$ the calculation (\ref{apr})
is rigorous since $\| \rho \|_{\mathcal E} \in C^1[0,T)$ and the
self-consistent potential satisfies $\Phi \in C^1([0,T);
C_b(\mathbb R^3))$.

In order to simplify the last term on the r.h.s. of (\ref{apr}) we
evaluate the trace in the eigenbasis of $\rho$ (cf. (\ref{fek})).
This gives
\begin{align*}
\frac{1}{2} \frac{d}{dt} \tr(\phi [\rho] \rho)= \frac{1}{2} \frac{d}{dt} \int_{\mathbb R^3}\phi (x)  n(x) dx.
\end{align*}
We now  proceed as in \cite{Ar}:  Integrating by parts several times and using the Poisson equation (\ref{pois}),
we obtain
\begin{align*}
\frac{1}{2} \frac{d}{dt} \tr(\phi [\rho] \rho) = & \
 \frac{1}{2} \frac{d}{dt} \int_{\mathbb R^3}|\nabla \phi[\rho]  (x) |^2dx =
 -  \int_{\mathbb R^3}\dot \phi[\rho]  (x) \Delta \phi[\rho]  (x) dx \\
= & \ \int_{\mathbb R^3}\dot \phi[\rho] (x) n[\rho]  (x) dx
=  \ \tr (\dot \phi [\rho] \rho) .
\end{align*}
Inserting this into (\ref{apr}), we get
\begin{align}\label{dte3}
 \frac{d}{dt} E^{tot}[\rho]= & \ \tr\left( -\frac{1}{2}\sqrt{-\Delta} \dot \rho \sqrt{-\Delta}
 + \frac{1}{2}|x| \dot \rho |x|
+\phi [\rho] \dot \rho \right) \nonumber \\
= & \ \tr\left( -\frac{1}{2}\sqrt{-\Delta} \mathcal L (\rho) \sqrt{-\Delta}    + \frac{1}{2}|x|  \mathcal L (\rho) |x|
+\phi [\rho] \mathcal L (\rho) \right).
\end{align}
In the following, we shall derive a differential inequality for $E^{tot}[\rho]$ from (\ref{dte3}).
This expression is now considerable easier to deal with, since
the self-consistent potential enters as if it was an additional external field
(note that the factor $1/2$ in front of $\phi [\rho ]$ has been eliminated).
\newpar
\emph{Step 2:} Similarly to the proof of lemma \ref{elem}, we introduce an energy-functional
\begin{align*}
E^{tot}_{k,l}[\rho ]:= E^{kin}_{k,l}[\rho ]+E_{k,l}^{ext}[\rho
]+\frac{1}{3}E^{sc} [\rho ], \quad k,l=1,2,3,
\end{align*}
where $E^{kin}_{k,l}$, $E_{k,l}^{ext}$ are defined as in (\ref{enfu}). Again, for all $\rho\in\mathcal D_\infty$, we have
\begin{align*}
E^ {tot}[\rho]=\sum_{k=1}^3 E^ {tot}_{k,k}[\rho]
\end{align*}
and, by a density argument, this carries over to $\rho\in\mathcal E$.
After some lengthy, but straightforward calculations (with extensive use of the cyclicity of the trace),
we get from (\ref{dte3}), the following equation:
\begin{align}\label{dte4}
\frac{d}{dt}\sum_{k=1}^3 E^ {tot}_{k,k}=
& \ \left(\frac{d}{dt}\sum_{k=1}^3 E^ {kin}_{k,k}-
\frac{i}{2}\sum_{k=1}^3 
\tr((\partial^2_k \phi[\rho])\rho+(\partial_k \phi[\rho])(\partial_k \rho))\right) \nonumber \\
& \ +  \frac{d}{dt}\sum_{k=1}^3 E^ {ext}_{k,k}
+ 2i \mu \sum_{k=1}^3 \tr (x_k \rho  (\partial _k \phi[\rho ])) \nonumber \\
& \ -i\sum_{k,l=1}^3 \sum_{j=1}^m \im (\overline{\alpha _{j,k}}\beta _{j,l})
\tr(x_k \rho (\partial _l \phi[\rho ] ))  \nonumber\\
& \ -i\sum_{k=1}^3 \sum_{j=1}^m \im(\overline{\gamma _j}\beta _{j,k})
\tr( \rho (\partial_k \phi [\rho ]) ).
\end{align}
Note that the first term of the r.h.s.~of (\ref{dte4}) -- in big
brackets -- equals the time derivative of $E_{k,k}^{kin}$ under
the \textit{linear} time-evolution. It is given by (\ref{dte1}).
On the other hand, one easily checks that the time derivative of
$E_{k,k}^{ext}$ under the nonlinear time-evolution is
equal to the linear one, hence given by (\ref{dte2}).
Since these kinetic and the external (potential) energy terms can
be treated (by interpolation arguments) as in the proof of lemma
\ref{elem}, it remains to estimate the last three terms on the
r.h.s. of (\ref{dte4}).
\newpar
Keep in mind, that we want to use a Gronwall lemma in the end. Hence,
we need to find appropriate \textit{linear} bounds for the r.h.s. of (\ref{dte4}).
(In the following we shall denote by $K$ positive, not necessarily equal, constants.)
\newpar
\emph{Step 3:} We first consider the term $ \tr( \rho (\partial_k \phi[\rho ]))$. 
In order to calculate the trace, we need to guarantee that $\rho
(\partial_k  \phi[\rho ])  \in \mathcal J_1$. Using the Sobolev
inequality we estimate for $\varphi \in L^2(\mathbb R^3)$:
$$
{\| (\sqrt{-\Delta} + I)^{-1} \varphi \| }_6 \leq \, K\, {\|(\sqrt{-\Delta} + I)^{-1} \varphi \| }_{H^1}
\leq  \, K\, {\| \varphi \| }_2,
$$
since ${\| (\sqrt{-\Delta} + I) \cdot \| }_2$ is an equivalent norm to ${\| \cdot \|}_{H^1}$. H\"older's inequality
and the bounds obtained in the proof of lemma \ref{lip} then imply
\begin{align*}
{\| (\partial_k  \phi[\rho ]) (\sqrt{-\Delta} + I)^{-1} \varphi \| }_2 \leq &
\, {\| \partial_k \phi [\rho ]   \|}_3 \ {\| (\sqrt{-\Delta} + I)^{-1} \varphi \| }_6\\
 \leq &  \, K {\| \partial_k \phi[\rho ]   \|}_3 \ {\| \varphi \| }_2.
\end{align*}
In other words, $(\partial_k \phi[\rho ]) (\sqrt{-\Delta} + I)^{-1} $ is a bounded operator on $L^2(\mathbb R^3)$ and we get
\begin{align*}
{\3norm \rho (\partial_k  \phi[\rho ])  \3norm}_1 \leq & \ {\3norm \, (\partial_k  \phi[\rho ]) 
(\sqrt{-\Delta} + I)^{-1}\3norm}_\infty  \, {\3norm \, (\sqrt{-\Delta} + I) \rho\,  \3norm}_1 \\
\leq  & \ K {\| \partial_k  \phi[\rho ] \|}_3 \, (E^{kin} [\rho]+{\3norm \rho \3norm}_1).
\end{align*}
Thus $\rho (\partial_k \phi[\rho ]) \in \mathcal J_1$, so we can calculate its trace in the eigenbasis of $\rho $ and estimate it:
\begin{align*}
|\tr( \rho (\partial_k \phi[\rho ] ))| = \left |\int_{\mathbb R^3}\partial_k \phi [\rho ] (x)  n[\rho ] (x) dx\right| \leq {\| \nabla \phi [\rho ]   \|}_2 \ {\| n[\rho ]   \|}_2 .
\end{align*}
The generalized Young inequality and the Lieb-Thirring inequality (\ref{lth1}) imply
\begin{align}\label{tr0}
{\| \nabla \phi [\rho ]   \|}_2 \leq K {\| n [\rho ]   \|}_{6/5}\leq K {\3norm \rho \3norm}_1^{3/4} E^{kin}[\rho ]  ^{1/4}.
\end{align}
Further, using again (\ref{lth1}), we have
$$
{\| n[\rho ]   \|}_{2}\leq K {\3norm \rho \3norm}_1^{1/4} E^{kin}[\rho ]  ^{3/4}.
$$
Hence, we obtain the following estimate:
\begin{align}
|\tr( \rho (\partial_k  \phi[\rho]) )| \leq  K {\3norm \rho \3norm}_1 \ E^{kin}[\rho ],
\end{align}
which is suitable for our purpose, due to the linear dependence on $E^{kin}[\rho]$.
\newpar
\emph{Step 4:} Next, we need to estimate the term
$$
\sum_{k,l=1}^3 \xi_{k,l}\tr(x_k \rho (\partial _l \phi[\rho ])),
$$
with the short-hand $\xi_{k,l}:=\im (\overline{\alpha _{j,k}}\beta _{j,l}).$ 
To guarantee that $x_k \rho  (\partial _l\phi[\rho ])\in \mathcal J_1$,
we only need to show $\sqrt{\rho} (\partial _l\phi[\rho]) \in \mathcal J_2$, since we already
know $x_k \sqrt{\rho}\in \mathcal J_2$. This can be done as in step 3 above by noting that
$\sqrt{\rho} (\sqrt{-\Delta} + I)\in \mathcal J_2$ and
$(\sqrt{-\Delta} + I)^{-1}\partial _l\phi[\rho] \in \mathcal B(L^2(\mathbb R^3))$.
\newpar
Hence, we can again calculate $\tr(x_k \rho (\partial _l \phi[\rho ]))$
in the eigenbasis of $\rho$:
\begin{align}\label{tr1}
\sum_{k,l=1}^3 \xi_{k,l}\tr(x_k \,\rho \,(\partial _l \phi[\rho ])) = & \
\sum_{k,l=1}^3 \xi_{k,l} \int_{\mathbb R^3} x_k\, \partial _l \phi[\rho ](x) \,n[\rho](x) dx
\nonumber \\
=& \ -\sum_{k,l,m=1}^3 \xi_{k,l} \int_{\mathbb R^3} x_k \,\partial _l \phi[\rho] (x) \,
\partial_{m,m}^2 \phi[\rho] (x) dx,
\end{align}
where we have used the Poisson equation (\ref{pois}) for the last equality. Integration
by parts gives
\begin{align}\label{tr2}
\sum_{k,l=1}^3 \xi_{k,l}\tr(x_k \,\rho \,(\partial _l \phi[\rho ])) =&
\sum_{k,l=1}^3 \xi_{k,l} \int_{\mathbb R^3} \partial _l \phi[\rho] (x)\,
\partial _k \phi[\rho] (x) dx
\nonumber\\
&+\sum_{k,l,m=1}^3 \xi_{k,l} \int_{\mathbb R^3} x_k \partial^2 _{l,m} \phi[\rho] (x)\,
\partial _m \phi[\rho] (x) dx .
\end{align}
Adding the equations (\ref{tr2}) and (\ref{tr1}) yields, after another integration by parts:
\begin{align}
& 2\sum_{k,l=1}^3 \xi_{k,l}\tr(x_k \,\rho \,(\partial _l \phi[\rho ]))\nonumber\\
= & \sum_{k,l=1}^3 \xi_{k,l} \int_{\mathbb R^3} \partial _l \phi\,
\partial _k \phi  \,dx
 \sum_{k,l,m=1}^3 \xi_{k,l} \int_{\mathbb R^3}
\left[x_k\, \partial _m \phi \, \partial_{l,m}^2 \phi
- x_k\, \partial _l \phi \, \partial_{m,m}^2 \phi \right] \,dx
\nonumber\\
=& \sum_{k,l=1}^3 \xi_{k,l} \int_{\mathbb R^3} \partial _l \phi\,
\partial _k \phi \,dx
\nonumber\\
& -\sum_{k,l,m=1}^3 \xi_{k,l} \int_{\mathbb R^3}
\left[\delta_{k,m}\, \partial_{l,m}^2 \phi + x_k\, \partial_{l,m,m}^3 \phi
-\delta_{k,l}\, \partial_{m,m}^2 \phi - x_k\, \partial_{l,m,m}^3 \phi \right]\,\phi \, dx
\nonumber\\
=& \ 2\sum_{k,l=1}^3 \xi_{k,l} \int_{\mathbb R^3} \partial_l \phi\,\partial _k \phi \,dx
-\sum_{k,m=1}^3 \xi_{k,k} \int_{\mathbb R^3} |\partial_m \phi|^2 \,dx,
\end{align}
where we write $\phi \equiv \phi [\rho]$ for simplicity and denote by $\delta_{k,l}$ the Kronecker symbol.
\newpar
Therefore we can estimate
$$
\left|\sum_{k,l=1}^3 \xi_{k,l}\tr(x_k \rho\, (\partial _l \phi[\rho ]))\right|
\leq K \,{\| \nabla \phi[\rho ] \|}_2^2,
$$
where $K$ depends on the coefficients $\xi_{k,l}$. 
Hence, using the same estimates as in (\ref{tr0}), we have
\begin{align*}
\left|\sum_{k,l=1}^3 \xi_{k,l}\tr(x_k \rho \, (\partial _l \phi[\rho ]))\right|
\leq & \ K \,{\3norm \rho \3norm}^{3/2}_1 \ E^{kin}[\rho ]^{1/2}\\
\leq & \ K  \,{\3norm \rho \3norm}_1 \left({\3norm \rho \3norm}_1+E^{kin}[\rho ]\right),
\end{align*}
which is the desired linear bound.
\newpar
The third term in (\ref{dte4}) can be treated analogously to the previous case.
\newpar
\emph{Step 5:}
The steps $1$-$4$, together with the estimates obtained in the
proof of lemma \ref{elem}, imply
\begin{equation}
 \label{noname}
 \frac{d}{dt} E^ {tot}[\rho(t)] \leq \  K E^ {tot}[\rho(t)],\quad 0\leq t < T,
\end{equation}
with some generic constant $K\geq 0$. Applying Gronwall's lemma then proves the assertion.
\newpar
Strictly speaking, all the calculations of steps $2-5$ first have to
be done for an approximating sequence $\{ \sigma_n \} \subseteq
\mathcal D_\infty$, such that $\sigma_n \stackrel{n\rightarrow
\infty}{\longrightarrow} \rho(t)$ in $\mathcal E$ for each fixed $t\in[0,T)$
(\cf the proof of theorem \ref{lth}). The estimate (\ref{noname})
then also holds for the limit $\rho(t)$ since the constant $K$ is
independent of $\{ \sigma_n \}$.
\newpar
\emph{Step 6:} So far we have proved (\ref{noname2}) for classical
solutions. By theorem \ref{thn1}(a) any mild solution (i.e.
$\tilde{\Phi}_t(\rho_0) \in C([0,T),\mathcal E)\,)$ can be approximated in
$\mathcal E$ (uniformly on $0 \le t \le T_1 < T$) by classical solutions.
Hence (\ref{noname2}) carries over to all initial conditions $\rho_0
\in \mathcal E$ with $\rho_0 \ge 0$.
\end{proof}
In view of (\ref{blow}), and since ${\| \rho(t) \|}_{\mathcal E} \le E^{tot}
[\rho(t)]$ we conclude from the above proposition that $T=\infty$ and
obtain our main result:
\begin{theorem}\label{thn2}
Let $\rho_0\in \mathcal E$, $d=3$ and $V_1\in L^\infty(\mathbb R^3)$ s.t.
$\nabla V_1\in L^q(\mathbb R^3)$, for some $3\leq q\leq \infty$:
\newpar
Then, the nonlinear evolution problem (\ref{nlqds}) admits a
unique mild solution, \ie it generates a nonlinear conservative
QDS: $\tilde{\Phi}_t(\rho_0)\in C([0,\infty), \mathcal E)$.
\end{theorem}


\section{Appendix: Proof of Lemma \ref{le}}

\renewcommand{\theequation}{A.\arabic{equation}}

Without loss of generality we can assume that $\rho$ is a
nonnegative operator. (Otherwise one can split $\rho $ into its
positive and negative part \cite{ReSi1} and prove the result
separately for each one.) Its eigenvalues are $\lambda_j \ge 0$
and the eigenvectors $\psi_j$ are orthonormal.
\newpar
\emph{Part (a):} For each $\rho \in \mathcal J_1^s$ with finite rank 
$N\in\mathbb N$ we shall show that the
approximation sequence $\{\sigma_n\} \subset \mathcal D_\infty$,
defined in (\ref{DDn}), satisfies $\sigma _n\rightarrow \rho $ in $\mathcal{J}_1$.
With the kernel of $\sigma_n \in \mathcal J_1^s$ as in (\ref{ker}), we get from \fer{conv1}, that 
$\sigma_n\rightarrow \rho$ in the strong operator topology.
Since we assumed that $\rho$ has finite rank, we
conclude from (\ref{est1}) that the trace norms converge, \ie
$$
\lim _{n\rightarrow \infty}{\3norm\sigma _n\3norm}_1= { \3norm\rho\3norm}_1.
$$
Then Gr\"umm's theorem (theorem~2.19 of \cite{Si})
implies $\sigma _n\rightarrow \rho $ in $\mathcal{J}_1$.
\newpar
\emph{Part (b):} The inclusion $\mathcal D(Z) \subset \mathcal D(\mathcal L)$ is already clear from proposition \ref{prop2}.
Thus it remains to show that for each $\sigma_n \in \mathcal D_n\subset \mathcal D_\infty$,
with some fixed $n\in \mathbb N$, we have $Z(\sigma_n)\in \mathcal J_1^s$: First, 
note that $Z(\sigma_n):=Y\sigma _n+\sigma _nY^*$ is a linear combination of the 
following terms (and their adjoints)
\begin{align}\label{terms}
x_k \sigma_n x_l, \ \partial_k \sigma_n \partial_l, \ \partial_k
\sigma_n x_l, \ x_k  x_l \sigma_n, \ \partial_k  \partial_l
\sigma_n, \ x_k  \partial_l \sigma_n, \ x_k \sigma_n, \ \partial_k
\sigma_n, \,
\end{align}
where $1\leq k,l\leq d$ and $\partial_k := \partial _{x_k}$.
(Indeed not all of this terms really appear in the expression of $Z$, but since the same argument for $\mathcal L$
is needed in the proof of theorem \ref{lth}, we shall consider this more general case.)\\
Since $\sigma_n$ has a representation given by $\sigma_n= M( \chi
_n)C(\varphi _n) \rho \ C(\varphi_n)M(\chi _n)$, for some $\rho
\in \mathcal J_1^s$, we have to prove that the operator
compositions $x^a \nabla^b M_n C_n$ are in $\mathcal B(L^2(\mathbb
R^d))$. Here the multi-indices $a, b \in \mathbb N_0^d$ are such
that $|a|+|b|\leq 2$. As an example we consider the operator $x_k
\partial _l$ and write for $f\in L^2 (\mathbb R^d)$:
\begin{align*}
(x_k \partial _l M_n C_n f)(x)= & \ x_k \partial _l (\chi_n(x)(\varphi_n \ast f)(x))\\
= & \ x_k [\partial _l\chi_n(x) (\varphi_n \ast f)(x) + \chi_n (x) (\partial _l \varphi_n \ast f)(x)].
\end{align*}
Since $\varphi,\chi\in C_0^\infty$ (see the proof of lemma \ref{le0}) we have that
$$
{\| x_k \partial_l M_n C_n f \|}_2 \leq K_{k,l,n} {\| f \|}_2
$$
and thus $x_k \partial_l M_n C_n \in \mathcal B(L^2(\mathbb
R^d))$.
Hence $x_k \partial_l \sigma_n = x_k \partial_l M_n
C_n \rho C_n M_n \in \mathcal J_1^s$. The other terms in
(\ref{terms}) can then be handled in a similar way.
\newpar
\emph{Part (c):} After the proof of part (a) it remains to show
that for all $\rho \in \mathcal J_1^s$ with $\mathcal L(\rho
)\in \mathcal J_1^s$, the following statement holds:
$$
\lim_{n\rightarrow \infty }{\3norm\mathcal L(\sigma _n) - \mathcal
L(\rho) \3norm}_1 = 0.
$$
To simplify the proof, it is sufficient to consider a ``model
operator" $\mathcal K(\rho)$, for which we choose $l=k=1$ in
(\ref{terms}) and further set all constants equal to one. This
simplification is possible since no cancellation occurs between
the individual terms of $\mathcal K(\rho)$. To simplify the
notation further, we shall from now on write $v:= x_1$,
$\partial:= \partial_{x_1}$. We choose $\mathcal K$ in the form
$$ 
\mathcal K(\rho)= \mathcal K_1(\rho) +\mathcal K_1(\rho)^*,
$$
where
$$
\mathcal K_1(\rho)=  v\rho v + \partial  \rho \partial +
\partial  \rho v + v^2 \rho + \partial ^2 \rho + v \partial \rho+
v \rho + \partial \rho.
$$
The general ($d$ - dimensional) case $\mathcal L(\rho) = -i [H,\rho]+A(\rho)$ described
above is then a straightforward extension.
The proof now follows again in several steps:
\newpar
\emph{Step 1:} We write
\begin{align*}
\mathcal K(\sigma _n)\equiv  & \  \mathcal K \big(M( \chi _n)C(\varphi _n) \rho \ C(\varphi _n)M(\chi _n)\big) \\
= & \ M( \chi _n)C(\varphi _n) \mathcal K(\rho) \ C(\varphi _n) M(\chi _n)  +R_n(\rho)+R_n(\rho)^* .
\end{align*}
Since $\mathcal K(\rho )\in\mathcal J_1^s$, we can decompose it into 
$\mathcal K(\rho) =\mathcal K_+(\rho)-\mathcal K_-(\rho)$, $\mathcal K_\pm(\rho)\geq 0$. 
Applying part (a) of this lemma then yields
\begin{equation*}
\lim_{n\rightarrow \infty}{\3norm \ M( \chi _n)C(\varphi _n) \mathcal K(\rho) \ M(\chi _n)C(\varphi _n)
- \mathcal K(\rho) \ \3norm}_1=0.
\end{equation*}
It remains to prove that $R_n(\rho) \rightarrow 0$ in $\mathcal
J_1$, as $n\rightarrow \infty$, which also implies $R_n(\rho)^*
\rightarrow 0$ in $\mathcal J_1$. For technical reasons (which
will become clear in step 3) we split this remainder term into two
parts: $R_n(\rho)  = R^1_n(\rho)  + R^2_n(\rho)$, and treat each
of them separately.
\newpar
\emph{Step 2:} After some lengthy calculations, $R^1_n (\rho)$
can be written as
\begin{align*}
\quad R_{ n}^1& \,(\rho) =
\ M (\partial \chi _n) C(\varphi _n)\, \rho \, C(\varphi _n) M(\partial \chi _n) \\
& + M(\partial ^2\chi _n) C(\varphi _n) \, \rho \, C(\varphi _n) M( \chi _n)  
+ M(\chi_n) C(v \varphi_n) \rho \, C(\varphi_n) M(\chi_n) \\
& + M(\partial \chi_n) C(\varphi_n) \, \rho \, C(\varphi_n) M(\chi_n) 
 - M(\chi _n) C(v^2\varphi _n) \, \rho  \, C(\varphi _n) M(\chi _n) \\
& - 2 \, M(\chi_n) C(v\varphi_n) \, \rho  \, C(v \varphi_n) M(\chi_n),
\end{align*}
where, on the level of the kernels, we have used several times the basic identity
$v(f\ast g)=vf\ast g + f\ast vg$.
Now we calculate for $f\in L^2(\mathbb R^d)$ (remember $v=x_1$)
\begin{align*}
(C(x_1\varphi _n) f)(x):&= \int_{\mathbb R^d} (x_1-y_1) \ \varphi _n(x-y) f(y)dy \\
&= \frac{1}{n} \int_{\mathbb R^d} n^{d+1}(x_1-y_1)\ \varphi (n(x-y)) f(y)dy = O\left(n^{-1}\right).
\end{align*}
Thus we have ${\3norm C(v\varphi _n) \3norm}_\infty = O\left(n^{-1}\right)$ and similarly we obtain
\begin{align*}
& {\3norm C(\varphi _n) \3norm}_\infty = {\3norm M(\chi _n)\3norm}_\infty = O(1), \\
& {\3norm M( \partial \chi _n)\3norm}_\infty=  O\left(n^{-1}\right), \\
& {\3norm C(v^2\varphi _n)\3norm}_\infty = {\3norm M(\partial ^2\chi _n)\3norm}_\infty=O\left(n^{-2}\right).
\end{align*}
With these relations we can estimate
\begin{align*}
{\3norm\ R^1_{n}(\rho) \ \3norm}_1 \leq & \  {\3norm\rho\3norm}_1 \,
{\3norm M(\chi _n)\3norm}_\infty^2  {\3normC(v\varphi _n)\3norm}_\infty^2 \\
&+\  {\3norm\rho\3norm}_1 \,{\3norm M(\chi _n)\3norm}_\infty^2
{\3normC(\varphi _n)\3norm}_\infty \,  {\3norm C(v^2\varphi _n)\3norm}_\infty \\
&+  \ {\3norm\rho \3norm}_1 \, {\3normC(\varphi _n)\3norm}_\infty^2  {\3norm M(\partial \chi _n )\3norm}_\infty^2 \\
&+ \ {\3norm\rho \3norm}_1 \, {\3normC(\varphi _n)\3norm}_\infty^2 
{\3norm M(\chi _n)\3norm}_\infty \, {\3norm M(\partial ^2\chi _n)\3norm}_\infty \\
&+  \ {\3norm\rho \3norm}_1 {\3norm C(\varphi _n)\3norm}_\infty {\3norm M(\chi _n)\3norm}_\infty^2 
{\3normC(v\varphi _n)\3norm}_\infty  \\
& +  \ {\3norm\rho \3norm}_1{\3normC(\varphi _n)\3norm}_\infty^2 {\3norm M(\chi _n)\3norm}_\infty \, 
{\3norm M(\partial \chi _n )\3norm}_\infty \\
=  & \ O\left(n^{-1}\right).
\end{align*}
Thus $R^1_n(\rho) \rightarrow 0$ uniformly in $\mathcal J_1$, as $n\rightarrow\infty$.
\newpar
\emph{Step 3:} Again a lengthy, but straightforward calculation
shows that the second part of the remainder can be written in the
form
\begin{align*}
\quad R_{n}^2 & \, (\rho ) =
 \ M(n \partial  \chi _n) C(\varphi _n) \, \rho \, C(\frac{\partial  \varphi _n}{n}) M(\chi _n)\\
& + M(\chi _n) C(\partial (v\varphi _n)) \, \rho \, C(\varphi _n) M(\chi _n)  
+ M(\chi_n) C(\frac{\partial \varphi_n}{n}) \, \rho \,C(\varphi_n) M(n\partial \chi_n) \\
& + M(n\partial \chi _n)C(\varphi _n)\, \rho \, C(\varphi _n) M(\frac{v}{n}\chi _n)
 + M(\chi _n)C(\frac{\partial \varphi _n}{n}) \, \rho \, C(nv\varphi _n)M(\chi _n) \\
& + M(v\partial \chi _n) C(\varphi _n) \, \rho \, C(\varphi _n) M(\chi _n) 
 + 2 \, M(n\partial \chi _n) C(\frac{\partial \varphi _n}{n})\, \rho C(\varphi _n) M(\chi _n)\\
& + M(\frac{v}{n} \chi_n) C(\varphi_n) \, \rho \, C(nv \varphi_n) M(\chi_n)
 + M(\chi _n) C(nv\varphi _n) \, \rho \, C(\varphi _n) M(\frac{v}{n}\chi_n)\\
& + 2 \, M(\frac{v}{n}\chi_n) C(nv\varphi_n) \, \rho \, C(\varphi_n) M(\chi_n).
\end{align*}
In contrast to step 2 these terms do not converge to zero uniformly in $\mathcal J_1$, 
hence we shall proceed differently: \\
As an example we consider the ninth term on the right hand side
and write
\begin{align*} M(\chi _n) C(nv\varphi _n) \rho \ C(\varphi _n)
M(\frac{v}{n}\chi _n) =
\ M(\chi _n) C(nv\varphi _n) \rho^N C(\varphi _n) M(\frac{v}{n}\chi _n) \   \\
+ \ M(\chi _n) C(nv\varphi _n) (\rho-\rho ^N) C(\varphi _n) M(\frac{v}{n}\chi _n),
\end{align*}
where $\rho ^N$ is the trace class operator $\rho $ ``cut" at finite rank $N\in\mathbb N$,
such that ${\3norm \ \rho -\rho ^N\3norm}_1\leq \varepsilon $, $\varepsilon \in \mathbb R_+$.
Direct calculations, similar to the one in step 2, imply
\begin{align} \label{ap1}
{\3norm C(nv\varphi _n)\3norm}_\infty \leq K,\ {\3norm M(n^{-1} v\chi _n) \3norm}_\infty \leq K, \ K\in \mathbb R,
\end{align}
with $K$ independent of $n\in \mathbb N$. Thus we can
estimate
\begin{equation}
\label{e61} \3normM(\chi _n) \ C(nv\varphi _n) (\rho-\rho ^N) \
C(\varphi _n) M(\frac{v}{n}\chi _n){\3norm}_1 \leq \varepsilon K^2.
\end{equation}
Define $\Pi$ to be the projector on $\ran(\rho ^N)$. Then
$\rho^N=\Pi\rho^N$ and
\begin{align}\label{ap2}
{\3norm C(nv\varphi _n) \rho ^N\3norm}_1 \leq  {\3norm C(nv\varphi _n) \Pi \3norm}_\infty  \ {\3norm \rho ^N\3norm}_1.
\end{align}
Now, since $\dim (\ran(\rho ^N)) <\infty$ and since strong convergence
equals uniform convergence on finite dimensional spaces \cite{ReSi1},
we get
\begin{equation}\label{ap3}
\lim_{n\rightarrow \infty } {\3norm C(nv\varphi _n) \Pi\3norm}_\infty =0.
\end{equation}
Combining (\ref{ap1}) - (\ref{ap3}) we thus have
\begin{equation}
\label{e62} \lim_{n\rightarrow \infty }\3normM(\chi _n) \ C(nv\varphi
_n) \rho^N \ C(\varphi _n) M(\frac{v}{n}\chi _n){\3norm}_1=0.
\end{equation}
Combining (\ref{e61}) and (\ref{e62}) shows that
\begin{equation*}
\3normM(\chi _n) \ C(nv\varphi _n) \rho \ C(\varphi _n)
M(\frac{v}{n}\chi _n){\3norm}_1
\end{equation*}
can be made arbitrarily small for $N$ sufficiently large.
All other terms appearing in the expression of $R_{n}^2$ can now be treated in the same way.\\
\newpar
In summary we have proved in steps 1 to 3 the assertion of the lemma.
\qed
\vskip 1 cm
\textbf{Acknowledgement:}
\newpar
This work has been supported by the Austrian Science Foundation
FWF through grant no. W8 and the \emph{Wittgenstein Award} 2000 of Peter Markowich.
Further support has been given by the European Union research network
\emph{HYKE}, by the DFG-project AR$277/3$-$2$ and by the
DFG-Graduiertenkolleg: \emph{Nichtlineare kontinuierliche Systeme
und deren Untersuchung mit numerischen, qualitativen und
experimentellen Methoden}.


\end{document}